\documentclass[fleqn,usenatbib]{mnras}

\usepackage{newtxtext,newtxmath}
\usepackage[T1]{fontenc}
\usepackage{underscore}
\usepackage{array}
\usepackage{multirow}
\DeclareRobustCommand{\VAN}[3]{#2}
\let\VANthebibliography\thebibliography
\def\thebibliography{\DeclareRobustCommand{\VAN}[3]{##3}\VANthebibliography}

\usepackage{graphicx}	% Including figure files
\usepackage{amsmath}	% Advanced maths commands
\newcommand{\gaia}{{\it Gaia}}

\newcommand{\coralie}{{CORALIE}}

\renewcommand{\arraystretch}{1.25}
\pdfoutput=1

\usepackage{xcolor}
\usepackage{soul}

\usepackage{longtable}
\usepackage{caption}

\title[EBLM XVIII]{The EBLM Project XVIII. 3D Obliquities of Five Low-Mass Eclipsing Binaries}

\author[Spejcher et al.]{
        Becca Spejcher$^{1*}$$^{\href{https://orcid.org/0009-0003-7036-2840}{\includegraphics[scale=0.5]{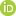}}}$,
        David V. Martin$^{1}$$^{\href{https://orcid.org/0000-0002-7595-6360}{\includegraphics[scale=0.5]{orcid.jpg}}}$,
        Jake Pandina$^{1}$$^{\href{https://orcid.org/0009-0005-1978-0320}{\includegraphics[scale=0.5]{orcid.jpg}}}$,
        Andy Zhang$^{1}$$^{\href{https://orcid.org/0000-0000-0000-0000}{\includegraphics[scale=0.5]{orcid.jpg}}}$,
        Max Ammons$^{1}$$^{\href{https://orcid.org/0009-0002-9966-5565}{\includegraphics[scale=0.5]{orcid.jpg}}}$,\newauthor
        Wata Tubthong$^{1}$$^{\href{https://orcid.org/0000-0002-7907-2634}{\includegraphics[scale=0.5]{orcid.jpg}}}$,
        Amaury Triaud$^{2}$$^{\href{https://orcid.org/0000-0002-5510-8751}{\includegraphics[scale=0.5]{orcid.jpg}}}$,
        Ritika Sethi$^{3}$$^{\href{https://orcid.org/0000-0002-6576-3346}{\includegraphics[scale=0.5]{orcid.jpg}}}$,
        Noah Vowell$^{4}$$^{\href{https://orcid.org/0000-0002-0701-4005}{\includegraphics[scale=0.5]{orcid.jpg}}}$,
        Adrian Barker$^{5}$$^{\href{https://orcid.org/0000-0003-4397-7332}{\includegraphics[scale=0.5]{orcid.jpg}}}$,\newauthor
        Pierre Maxted$^{6}$$^{\href{https://orcid.org/0000-0003-3794-1317}{\includegraphics[scale=0.5]{orcid.jpg}}}$,
        Alison Duck$^{8,7}$$^{\href{https://orcid.org/0000-0002-4531-6899}{\includegraphics[scale=0.5]{orcid.jpg}}}$,
        Shelby Summers$^{7}$$^{\href{https://orcid.org/0009-0006-8371-7227}{\includegraphics[scale=0.5]{orcid.jpg}}}$,
        François Bouchy$^{9}$$^{\href{https://orcid.org/0000-0002-7613-393X}{\includegraphics[scale=0.5]{orcid.jpg}}}$,
        Monika Lendl$^{9}$$^{\href{https://orcid.org/0000-0001-9699-1459}{\includegraphics[scale=0.5]{orcid.jpg}}}$,\newauthor
        Maxime Marmier$^{9}$$^{\href{https://orcid.org/0000-0001-5630-1396}{\includegraphics[scale=0.5]{orcid.jpg}}}$, 
        Vincent Megevand$^{9}$,
        Francesco Pepe$^{9}$$^{\href{https://orcid.org/0000-0002-9815-773X}{\includegraphics[scale=0.5]{orcid.jpg}}}$,
        Malte Tewes$^{10}$$^{\href{https://orcid.org/0000-0002-1155-8689}{\includegraphics[scale=0.5]{orcid.jpg}}}$,
        Stéphane Udry$^{9}$$^{\href{https://orcid.org/0000-0001-7576-6236}{\includegraphics[scale=0.5]{orcid.jpg}}}$\\
$^{1}$Department of Physics and Astronomy, Tufts University, Medford, MA 02155, USA\\
$^{2}$School of Physics and Astronomy, University of Birmingham, Edgbaston, Birmingham B15 2TT, UK\\
$^{3}$Department of Physics, Massachusetts Institute of Technology, Cambridge, MA 02139, USA\\
$^{4}$Department of Physics and Astronomy, Michigan State University, East Lansing, MI 48824, USA\\
$^{5}$School of Mathematics, University of Leeds, Leeds LS2 9JT, UK \\
$^{6}$Astrophysics Group, Keele University, Keele, Staffordshire, ST5 5BG, UK\\
$^{7}$Department of Astronomy, The Ohio State University, Columbus, OH 43210, USA\\
$^{8}$Jet Propulsion Laboratory, California Institute of Technology, 4800 Oak Grove Drive, Pasadena, CA 91109, USA\\
$^{9}$Observatoire Astronomique de l’Université de Genève, Chemin Pegasi 51, CH-1290 Versoix, Switzerland\\
$^{10}$ Universität Bonn, Argelander-Institut für Astronomie, Auf dem Hügel 71, 53121 Bonn, Germany
}

\date{Under Review (first submission November 2025, second submission May 2026)}

\pubyear{2026}

\begin{document}
\label{firstpage}
\pagerange{\pageref{firstpage}--\pageref{lastpage}}
\maketitle

% Abstract of the paper
\begin{abstract}
The formation of tight stellar binaries remains an unsolved problem. There is too much angular momentum in a collapsing and fragmenting protostellar cloud to form a stellar binary in situ with a separation less than an AU, yet thousands of these short-period binaries have been discovered. One indication of a binary's formation is the angle between the stellar spin and orbital axes --- its obliquity. The classical method for determining projected stellar obliquities is the Rossiter-McLaughlin effect. This has been applied to 132 hot Jupiters, but only a handful of stellar binaries. Of the binary systems with measured projected obliquities, even fewer have measured 3D obliquities. In this paper, we add five more short-period binary 3D obliquity measurements to the sample previously consisting of one system. We present Rossiter-McLaughlin measurements for EBLM J0239-20, EBLM J0941-31, EBLM J1037-25, EBLM J1141-37, and EBLM J2025-45. These systems consist of an M-dwarf eclipsing an F/G type primary. We combine CORALIE and HARPS spectroscopy with TESS photometry of primary and secondary eclipses. We show that even though the sky-projected obliquities seem to be aligned, there is modest but non-zero spin-orbit misalignment ($\psi$ between 5 and 20$^{\circ}$). Our primary stars straddle the Kraft break at $\sim 6250K$. We derive the M-dwarf masses and radii to precisions better than 3\%. With the exception of EBLM J0941-31, each system has an inflated radius, exceeding stellar model predictions by more than 5$\sigma$.

\end{abstract}

% Select between one and six entries from the list of approved keywords.
% Don't make up new ones.
\begin{keywords}
stars: binaries-eclipsing, low-mass, fundamental parameters; techniques: photometric, spectroscopic
\end{keywords}

%%%%%%%%%%%%%%%%%%%%%%%%%%%%%%%%%%%%%%%%%%%%%%%%%%

%%%%%%%%%%%%%%%%% BODY OF PAPER %%%%%%%%%%%%%%%%%%

\section{Introduction}

The origin of tight stellar binaries continues to perplex astronomers. Collapsing and fragmenting clouds have too much angular momentum to place two stars on a separation of a fraction of an AU (orbital periods of $\sim$days, \citealt{Tohline2002,Moe2018}), yet thousands of short-period main sequence binaries have been discovered \citep{Raghavan2010,Prsa2011,Jayasinghe2018,Prsa2022,Mowlavi2023}. One piece of the formation puzzle is the spin-orbit obliquity of the binary. This can be measured using the Rossiter-McLaughlin (RM) effect \citep{Rossiter1924,McLaughlin1924,Queloz2000b,Triaud2017}. This spectroscopic effect can be seen when an object, like a planet or a secondary star, transits the primary star. This technique has been primarily applied to hot Jupiters, with over 100 RM measurements to date, revealing diverse orbits: coplanar, misaligned, polar, and even retrograde \citep{Winn2005,Fabrycky2007,Triaud2010,Albrecht2012,Dawson2018,Rice2022, Rossi2025}. Brown dwarfs, which straddle the line between hot Jupiter and M dwarf have a sky-projected obliquity of $|\lambda| < 10^{\circ}$ in 10 out of 12 cases \citep{Carmichael2025, Vowell2025b}. Only a handful of RM effects have been measured for stellar binaries. Furthermore, the RM effect alone only provides a \textit{projected} spin-orbit obliquity. An independent measurement of the stellar rotation rate is needed to determine the \textit{true 3D} obliquity. The majority of obliquity measurements in the literature are only projected, not true obliquities.

%A misalignment might speak to a dynamical history (e.g. scattering or Kozai-Lidov cycles), whereas an alignment might imply a more placid history (e.g. in situ formation or smooth disc migration). To complicate the matter, star-planet tidal interactions tend to re-align the planet's orbit over time, potentially washing away the original formation signature. 

In 2010, we created the EBLM (Eclipsing Binary Low Mass) survey (see review in \citealt{Maxted2023}). The sample consists of M-dwarfs ($\approx 0.08 - 0.6 M_\odot$) eclipsing more massive F/G/K dwarfs. These systems were discovered as ``false positives'' from the WASP transiting hot Jupiter survey. There are multiple goals for the EBLM project, one of which is increasing the number of precise obliquities of eclipsing binaries. So far, the EBLM project has published obliquities for WASP-30 (an eclipsing brown dwarf) \& EBLM J1219-39 \citep{Triaud2013}, EBLM J0218-31 \citep{Gill2019}, the circumbinary planet host EBLM J0608-59 (also known as TOI-1338, BEBOP-1) \citep{hodzic2020}, and EBLM J0021-16 \citep{Spejcher2025b}.

In this paper, we present RM measurements for five more EBLM systems: EBLM J0239-20, EBLM J0941-31, EBLM J1037-25, EBLM J1141-37, and EBLM J2025-45. The radial velocities both in and out-of-eclipse were observed with the CORALIE spectrograph. Additionally, EBLM J0941-31, EBLM J1037-25, and EBLM J1141-37 have out-of-eclipse radial velocities that were observed with the HARPS spectrograph. All five systems were observed by TESS, where photometry shows primary and secondary eclipses. For all five binaries, TESS photometry also reveals star spot modulation, from which we can derive the primary star's rotation rate and hence determine a true 3D spin-orbit obliquity. Compared to 132 hot Jupiter obliquity measurements, only eight eclipsing binaries have obliquity measurements. Even less-so, only one short-period eclipsing binary has had a true 3D obliquity measurement \citep{Spejcher2025b}. Our sample adds five more sky-projected and true 3D obliquity measurements for short period eclipsing binaries, essentially quintupling the number of binaries with true 3D obliquity measurements.

We find that out of five targets, four have sky-projected obliquities of $|\lambda|<5^\circ$ and one has a sky-projected obliquity of $|\lambda|=11.7^\circ$. The true 3D obliquities are larger, ranging from $\psi \approx 15^\circ$ to $\psi \approx 25^\circ$.

Our paper is organized into the following sections: in Sect. \ref{section:observations} we describe the spectroscopic and photometric observations; in Sect. \ref{section:methods} we discuss our data-fitting methods; in Sect. \ref{section:results} we discuss our results and in Sect. \ref{section:discussion} we discuss implications of our results with respect to tidal physics, other measured obliquities in the literature, and M-dwarf radius inflation, followed by our conclusion. 
%%%%%%%%%%%%%%%%%%%%%%%%%%%%%%%%%%%%%%%%%%%%%%%%%%%%%
%               Observations                        %
%%%%%%%%%%%%%%%%%%%%%%%%%%%%%%%%%%%%%%%%%%%%%%%%%%%%%
\section{Observations}\label{section:observations}

Each of the five targets was first observed by the WASP transiting hot Jupiter survey. They were candidate hot Jupiters, with a ``transit'' depth on the order of a a couple of per cent. Follow-up radial velocity observations with \coralie ~revealed that each of these systems was a single-lined spectroscopic eclipsing binary, with an M-dwarf secondary. 

\subsection{EBLM J0239-20}

EBLM J0239-20 has an orbital period of $P_{\rm orb} \sim 2.78$ days. First published by \citet{Triaud2017}, the system was flagged for needing a linear drift added to the single Keplerian orbit model in order to fit the RVs, which is indicative of the system having a tertiary companion. This is in spite of a small  \gaia ~re-normalized unit weight error (RUWE) of 0.94 (\ref{Observation_table}), where a value greater than 1.5 would be a strong astrometric indicator of a tertiary companion. The \gaia  ~RUWE, however is neither a necessary nor sufficient condition for an additional star; the RVs are a more reliability indicator of a tertiary star here.
\subsection{EBLM J0941-31}\label{EBLM J0941-31}

EBLM J0941-31 has an orbital period of $P_{\rm orb} \sim 5.55$ days. First published by \citet{Triaud2017}, the RVs of the system were fit using a single Keplerian orbit. Although no linear drift was fit in \citet{Triaud2017}, the \gaia~RUWE is 1.26, so there might be a tertiary star. Furthermore, our fit of EBLM J0941-31 does include a linear drift fit. For both this system and EBLM J0239-20, \gaia~DR4, which is coming mid 2026, will help further characterize possible third bodies in these systems. EBLM J0941-31 is one of three targets in our sample that was placed in the BEBOP (Binaries Escorted By Orbiting Planets) radial velocity survey for circumbinary planets, which resulted in it receiving more spectroscopy from both CORALIE and the higher resolution HARPS spectrograph \citep{Martin2019}. Although the system did not show evidence for a circumbinary planet after further investigation, this paper uses the HARPS data to more tightly constrain the parameters of our RV fits.

\subsection{EBLM J1037-25}

EBLM J1037-25 has an orbital period of $P_{\rm orb} \sim 4.94$ days. It was first observed in data release two of the radial velocity experiment (RAVE), released in July 2008 \citep{Zwitter2008}. The system was then flagged as a hot Jupiter candidate by WASP and then confirmed to be a single-line spectroscopic eclipsing binary in \citet{Triaud2017}. EBLM J1037-25 was also part of the BEBOP circumbinary planet survey \citep{Martin2019}.

\subsection{EBLM J1141-37}

EBLM J1141-37 has an orbital period of $P_{\rm orb} \sim 5.15$ days. It was first published in 1982 in the \textit{Michigan Catalogue of Two-dimensional Spectral Types for the HD stars} as HD 101579 \citep{Houk1982}. The system was then observed in the X-ray as a candidate member of the TW Hya association \citep{Tachihara2003}. The system was finally published as an EBLM system in \citet{Triaud2017}. EBLM J1141-37 was also part of the BEBOP circumbinary planet survey \citep{Martin2019}.

\subsection{EBLM J2025-45}

EBLM J2025-45 has an orbital period of $P_{\rm orb} \sim 6.19$ days. It was first published in \citet{Triaud2017}, where it was flagged as an active system based on variation of the bisector of the CCF.

\begin{table*}
\caption{Summary of observational properties.}             % title of Table
\label{Observation_table}      % is used to refer this table in the text
\centering                                    % used for centering table
%\resizebox{0.95\linewidth}{!}{
\begin{tabular}{m{2cm} m{2.65cm} m{2.65cm} m{2.65cm} m{2.65cm} m{2.65cm}}  

\hline\hline                        % inserts double horizontal lines
 
 & EBLM J0239-20 &  EBLM J0941-31 & EBLM J1037-25 & EBLM J1141-37 & EBLM J2025-45\\
 
\hline 
 TIC & 64108432 & 25776767 & 187930369 & 454013359 & 369984032\\
 {\it TESS} Sectors & 4, 31 & 8, 9, 35, 62, 89 & 9, 36, 63, 89 & 10, 36, 37, 63, 90 & 27, 67\\ 
 2MASS & J02392928-2002238 & J09411676-3149102 & J10370694-2534177 & J11411219-3747297 & J20252547-4549451\\
 \gaia ~DR3 ID & 5127259153974964480 & 5440085875822885888 & 5470783042175620992 & 5385134144388660352 & 6669738865254236928\\\\
 
RA ($\alpha$) [BCRS]
& $02^{\rm h}39'29.29''$ & $09^{\rm h}41'16.76''$ & $10^{\rm h}37'06.94''$ & $11^{\rm h}41'12.19''$ & $20^{\rm h}25'25.48''$\\
& $39.872184^{\circ}$ & $145.319703^{\circ}$ & $159.278617^{\circ}$ & $175.300470^{\circ}$ & $306.356421^{\circ}$\\

Dec ($\delta$) [BCRS]
& $-20^{\circ}02' 23.99''$ & $-31^{\circ}49'10.31''$ & $-25^{\circ}34'17.73''$ & $-37^{\circ}47'29.62''$ & $-45^{\circ}49'45.16''$\\
& $-20.040050^{\circ}$ & $-31.819494^{\circ}$ & $-25.571653^{\circ}$ & $-37.791613^{\circ}$ & $-45.829560^{\circ}$\\ \\

CORALIE Obs \# & 52 & 37 & 38 & 52 & 64\\

CORALIE Obs $\Delta$ t [days] & 1201 & 1839 & 1267 & 2176 & 1991\\
CORALIE Obs Date [UTC] & August 2009 - December 2012 & March 2008 - March 2013 & December 2009 - June 2013 & August 2008 - July 2014 & May 2009 - November 2014\\
CORALIE RM Obs Date [UTC] & November 11, 2009; December 19, 2009 & February 14, 2010 & March 10, 2010 & April 24-25, 2010 & September 7, 2010\\ \\
HARPS Obs Date [UTC] & & May 2018 - May 2022 & May 2018 - March 2021 & April 2018 - December 2022 & \\ \\

$TESS$-mag & $10.1356 \pm 0.0081$ & $10.638 \pm 0.006$ & $9.6475 \pm 0.0062$ & $8.995 \pm 0.006$ & $10.3789 \pm 0.0061$\\

$B$-mag & $11.34 \pm 0.07$ & $11.62 \pm 0.06$ & $10.66 \pm 0.01$ & $10.13 \pm 0.04$ & $11.87 \pm 0.12$\\

$G$-mag & $10.568 \pm 0.003$ & $11.039 \pm 0.003$ & $10.020 \pm 0.003$ & $9.397 \pm 0.003$ & $10.892 \pm 0.003$\\

$V$-mag & $10.63 \pm 0.05$ & $11.08 \pm 0.06$ & $10.16 \pm 0.04$ & $9.57 \pm 0.03$ & $11.15 \pm 0.09$\\

$G_{\rm BP} - G_{\rm RP}$ & 0.813 & 0.756 & 0.694 & 0.743 & 0.967\\

Parallax [mas] & $3.867 \pm 0.0156$ & $2.670 \pm 0.022$ & $4.296 \pm 0.018$ & $5.654 \pm 0.015$ & $6.310 \pm 0.024$\\ 
% Distance [pc] & 254.9135 &   & 231.2979 & 177.2652 & 157.1909\\ 
\gaia ~RUWE & 0.94010013 & 1.2649149 & 0.9237525 & 0.99194455 & 1.2546798\\

\hline

\end{tabular}
\end{table*} 

\subsection{Radial Velocity Spectroscopy}

The observational properties of the five EBLM systems are summarized in Table~\ref{Observation_table}. The systems were first discovered photometrically by the Wide Angle Search for Planets (WASP) and were labeled as hot Jupiter candidates. Follow-up radial velocities (RVs, Fig. \ref{fig:RVs}) were taken using the \coralie ~spectrograph, which is a fiber-fed \'{e}chelle spectrograph installed on the 1.2-m Leonard Euler telescope at the ESO La Silla Observatory, and has a resolving power of R = 50,000\,--\,60,000 \citep{Queloz2000a}. \coralie ~has a wavelength range of $390-680$ nm. It was quickly revealed that all five systems were not planets, but rather low mass eclipsing binaries, placing them into the EBLM program. Continued RVs were taken between March 2008 and November 2014 to map out the Keplerian orbit of the binary. In addition, each system received a sequence of spectroscopic measurements during the primary eclipse, yielding the RM effect. For EBLM J0239-20, the RM effect was observed on two separate occasions, owing to the relatively low signal to noise ratio of the effect. All RV data were processed using the standard \coralie ~data reduction pipeline \citep{Baranne1996} using the cross-correlation method.

Additionally, three out of the five systems were observed with the High Accuracy Radial velocity Planet Searcher \citep[HARPS;][]{Mayor2003} spectrograph on the 3.6m telescope at the ESO La Silla observatory as part of the Binaries Escorted By Orbiting Planets (BEBOP) survey for circumbinary planets. HARPS is a fiber-fed \'{e}chelle spectrograph that uses a laser frequency comb to obtain an accuracy on the order of 1m/s. HARPS has a resolving power of R = 115,000 and covers the 378-691 nm wavelength range. The targets were observed at different times from April 2018 to December 2022; however, during this time, no eclipses were observed. The exact observation dates of each system both for \coralie ~and HARPS are provided in Table \ref{Observation_table}.

\subsection{TESS Photometry}\label{subsec:photometry}

Each system has been observed by TESS for at least two sectors. The sectors during which they were observed are listed in Table~\ref{Observation_table}. For all of the systems, we downloaded the PDCSAP (Pre-search Data Conditioned Simple Aperture Photometry) light curves via \textsc{lightkurve} \citep{lightkurve2018}. We downloaded the 120-second exposure time data that had been reduced through the SPOC (Science Processing Operations Center) pipeline. We chose to use the PDCSAP flux because it accounts for some instrumental effects, cosmic ray mitigation, and is corrected for dilution from nearby stars. We note that for our recent paper on the RM of EBLM J0021-16 \citep{Spejcher2025b} we used the SAP flux because the PDCSAP flux cut out a couple of eclipses and also had a detrending artefact that could be confused as the transit of a circumbinary planet. Such issues were not present in the five targets in this paper.

\section{Methods}\label{section:methods}

\subsection{Primary Star Parameters Using \textsc{Exofastv2}}

The radius and effective temperature of the primary stars were constrained using \textsc{Exofastv2} \citep{Eastman2019} to fit a spectral energy distribution (SED) in combination with the MESA Isochrone and Stellar Tracks (MIST) model \citep{Dotter2016}. This approach only applies the SED fitting and the MIST model to the FGK primary stars and not the eclipsing M-dwarf, whose properties are derived from transit fitting and the radial velocity measurements.  The bolometric flux ratio between the primary and secondary stars is 

\begin{equation}
    \frac{F_{\rm B}}{F_{\rm A}} = \left(\frac{R_{\rm B}}{R_{\rm A}}\right)^2\left(\frac{T_{\rm eff, B}}{T_{\rm eff, A}}\right)^4,
\end{equation}
which for our stars ranges between 0.003 for EBLM J1141-37 and 0.0006 for EBLM J02339-20. Therefore, a single-star SED fit is reasonable in all cases for our sample since the eclipsing M-dwarf contributes well under $1\% $ of the total flux.

The observed broadband photometric data are taken from \gaia ~DR3 \citep{Gaiadr3}, 2MASS \citep{cutri_2003}, and WISE \citep{cutri_2014}. We adopted priors on the primary star effective temperature and metallicity from \citet{Freckelton2024} as available for EBLM J0941-31, EBLM J1037-25, and EBLM J1141-37. We did not impose a stellar effective temperature or metallicity prior on EBLM J0239-20 and EBLM J1141-37. As the targets in our sample are a single-lined spectroscopic binaries, we cannot directly measure both stellar masses in a model-independent way. Thus, we use \textsc{Exofastv2} to fit the MIST models \citep{Dotter2016} to estimate the primary star masses, as was done in previous EBLM work such as \citet{Duck2023}.

% \subsection{Data Processing} \label{Processing}
 \subsection{Light Curve Detrending} 
 We detrend the TESS light curves using the \textsc{Wotan} \citep{Hippke2019}
 Python package to remove any non-eclipse variability. This includes the effects of spot modulation on the primary star, gyroscopic movements from TESS, and scattered light by the Earth. In order to keep the primary and secondary eclipses in the lightcurves of each system, the \textsc{wotan} transit mask function was applied to the data. We then applied a spline with robust Huber-estimator and a break tolerance of 0.5. The window lengths chosen for our targets were calculated to be roughly three times their eclipse duration. The need for a more robust method of detrending was due to significant star spot modulation. For EBLM J0941-31, we applied a Tukey's biweight filter with a window length of 0.6. This filter was necessary for this system because although stellar variability is present, it was not enough to require a more robust method of detrending. 
 
 % We apply a Tukey's biweight filter with the window length of 0.7 days to ensure that the eclipse depths and shapes are preserved while other activity is removed. The raw and detrended light curves are shown in Fig. \ref{fig: Eclipses and Light curves}.

\subsection{Spectroscopy and Photometry Model Fit}

\subsubsection{\textsc{Pymc3} and \textsc{Exoplanet}}\label{subsubsec:Fitting RVs}

To determine the physical properties and orbital parameters of both stars in the binary, we use the \textsc{Exoplanet} \citep{exoplanet:joss} package to model the eclipse light curves and RV profile, and employ the MCMC algorithm to derive the posterior of each properties. \textsc{Exoplanet} contains a suite of functionality that includes solving Kepler's equation, \citet{Mandel2002} light curves, quadratic limb darkening parameterization, and specialized impact parameter priors. Although originally developed for modeling stars with exoplanets, \citet{exoplanet:joss} is applicable to eclipsing binaries and has been used in past studies including \citet{Pass2023}, \citet{Martin2024Outliers,Martin2024CMDra}, and \citet{Spejcher2025b}.

The \textsc{Pymc3} package \citep{pymc3} is then used in conjunction with \textsc{Exoplanet} to determine the best-fit parameters via the MCMC method. We use the built-in optimizer to find the \textit{maximum a posteriori (MAP)} set of parameters and use it as the starting point of the MCMC sampling. The optimizer utilizes the Limited-memory Broyden–Fletcher–Goldfarb–Shanno (L-BFGS-B; \citealt{LMBFGS}) algorithm. The default fitting algorithm is the No-U-Turn Sampler (NUTS; \citealt{NUTS}), which we use for all of our fits.

\subsubsection{Fitting Keplerian RVs} \label{RV Fitting}

We first perform a fit solely based on the Keplerian RV model, i.e. in-eclipse RVs were ignored. For EBLM J0239-20 and EBLM J2025-45, this follows the same technique used in \citet{Spejcher2025b} where we only have \coralie ~RV data. For EBLM J0941-31, EBLM J1037-25, and EBLM J1141-37, we adjusted our methods to simultaneously for the RVs from \coralie ~and HARPS. Both data sets will have the same parameters for the RV semi-amplitude ($K$), orbital period ($P_{\rm orb}$), time of eclipse ($T_{\rm 0}$), eccentricity ($e$), and the argument of periastron ($\omega$), which are fit as if there is only one set of RV data. However, because the \coralie ~data were taken at different times than the HARPS data and come from different instruments, each data set will have a different systemic radial velocity ($\gamma$), which we have to fit for separately and subtract from each respective data set. $T_{\rm 0}$, $\gamma$, $\log(K)$, and $\log(P_{\rm orb})$ were sampled using a normal distribution with a mean around their priors and a reasonable standard deviation. We sample over $\log(K)$ and $\log(P_{\rm orb})$ instead of the semi-amplitude and period to more efficiently sample the parameter space. For $e$ and $\omega$, rather than sampling for these two parameters directly, we sample $h=\sqrt{e}\sin\omega$ and $k=\sqrt{e}\cos\omega$ to avoid introducing a biased estimate for $e$.

In order to obtain a fit of EBLM J0239-20 and EBLM J0941-31 with minimal residuals, a linear drift parameter was needed 
($\dot{\gamma}$). This parameter is a strong indicator of the presence of a distant third star in these systems.

We use \textsc{exoplanet} to model the radial velocity of a Keplerian orbit. We then optimize the model parameters to find the MAP model and perform an MCMC with \textsc{Pymc3}. The resulting RV parameters from this model are then used as priors for the joint RV and photometry fit.

\subsubsection{Joint RV and Photometry Fit}

After getting the best-fit parameters from the Keplerian RV fit, we create a joint fit for the RV and photometry data sets. The posterior values from the RV fit are taken as priors for the parameters mentioned in Section \ref{RV Fitting}. Additional parameters in the joint fit include the mass and radius of the primary star ($M_{\rm A}$, $R_{\rm A}$), the mass and radius of the secondary star ($M_{\rm B}$, $R_{\rm B}$), impact parameter ($b$), and quadratic limb darkening coefficients of both stars. To fit for the mass and radius of the primary star for EBLM J0239-20 and EBLM J0941-39, we use values from \citet{Swayne2024} as our priors. For EBLM J1037-25 and EBLM J1141-37, we use values from \citet{Freckelton2024} as our prior primary mass and radius values. For EBLM J2025-45, we used values from \citet{Triaud2017} as our primary mass and radius priors. The secondary star's mass and radius are calculated using the primary star's parameters, multiplied by the mass ratio ($q=M_{\rm A}/M_{\rm B}$) and radius ratio ($R_{\rm B}/R_{\rm A}$) respectively. The mass ratio is sampled on a logarithmic scale, while the radius ratio is sampled on a linear scale. We sample the impact parameter ($b$) using the built-in \textsc{exoplanet} function \textit{xo.distributions.ImpactParameter}, which samples from a uniform distribution between zero to $1+R_{\rm B}/R_{\rm A}$ \citep{exoplanet:joss}. 

The primary and secondary eclipse light curves are modeled using \textit{xo.SecondaryEclipseLightCurve} from \textsc{Starry} \citep{Luger2019}. The quadratic limb darkening coefficients are fitted using \textit{xo.QuadLimbDark} function. 

Similar to the RV fit, we optimize the starting parameters for MAP and then perform the MCMC sampling. We then use the posteriors from our joint fit as the priors to fit for the RM anomaly.

% period, time of eclipse, and $h=e\sin\omega$ and $k=e\cos\omega$. 

\begin{figure*}
  \centering
  \begin{tabular}{cc}
    \includegraphics[width=0.49\textwidth,trim={3cm 1cm 4cm 1cm},clip]{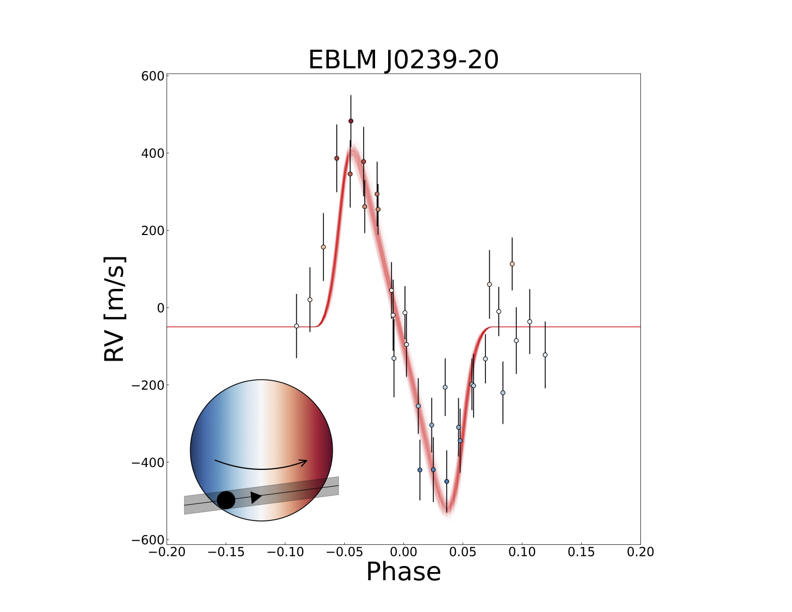} &
    \includegraphics[width=0.49\textwidth,trim={3cm 1cm 4cm 1cm},clip]{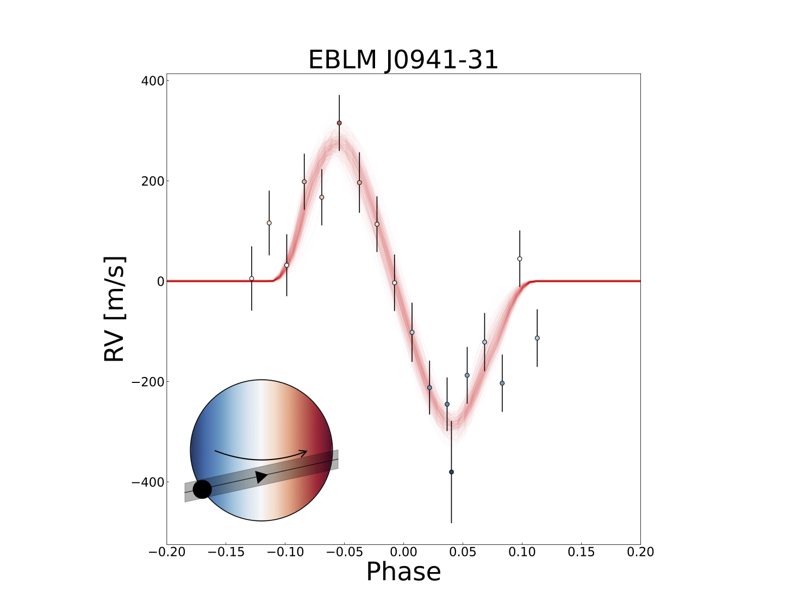} \\
    
    \includegraphics[width=0.49\textwidth,trim={3cm 1cm 4cm 1cm},clip]{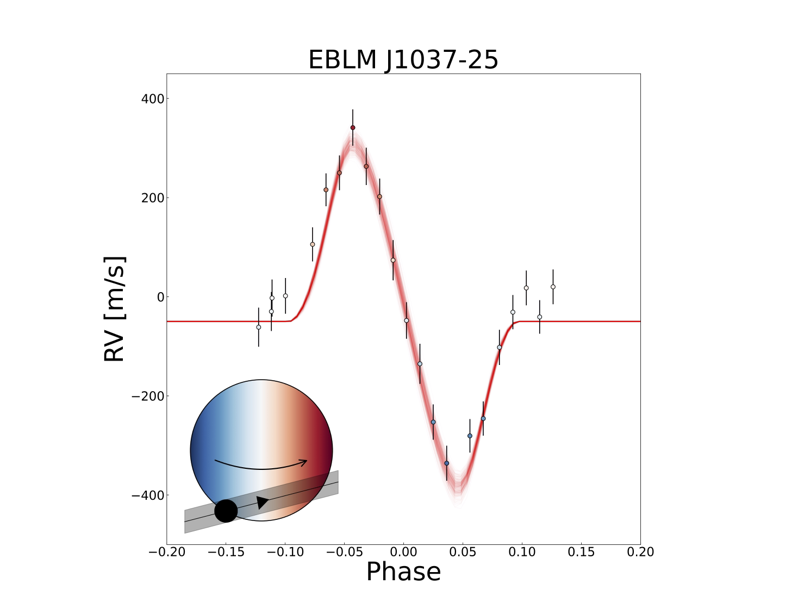} &
    \includegraphics[width=0.49\textwidth,trim={3cm 1cm 4cm 1cm},clip]{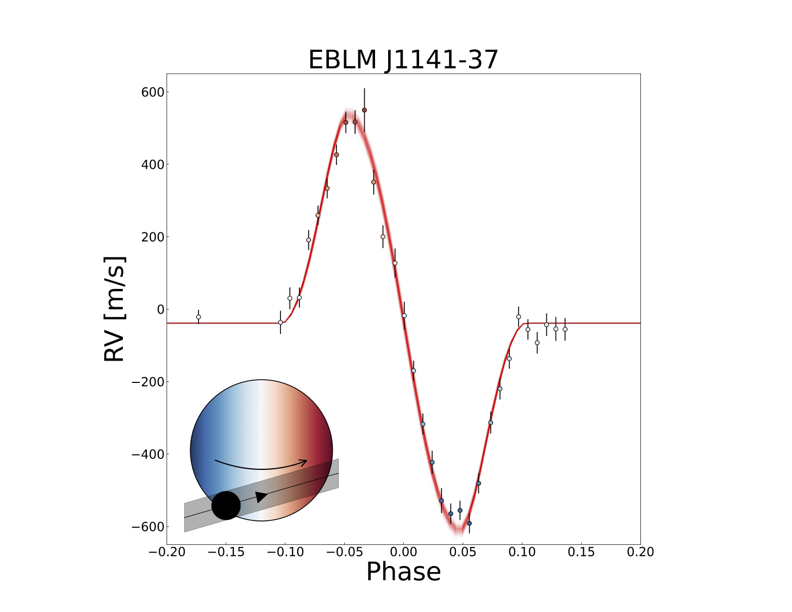} \\
    
    \includegraphics[width=0.49\textwidth,trim={3cm 1cm 4cm 1cm},clip]{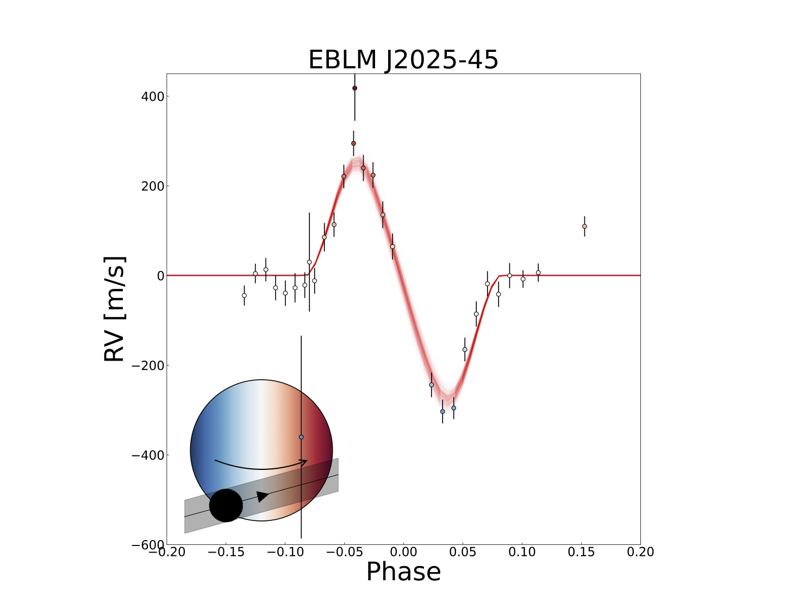} &
    \parbox[b][0.45\textwidth][c]{0.49\textwidth}{%
      % \centering
      \caption{Fit of the RM effect for each system. The points are colored based on the Doppler shift occurring during the primary eclipse. The diagram in the bottom left corner gives a visual representation of the orbit base on the impact parameter and true obliquity values from the MCMC fit. While the fit of the RM effect only gives us the sky-projected obliquity, we could find the true obliquity using the rotation period of each system. As the secondary star is blocking the blue light from the primary star, the radial velocity points get red-shifted, and as the red light is being blocked, the radial velocity points get blue-shifted.}\label{fig:RM Example Graph}}
  \end{tabular}

\end{figure*}

\begin{figure}
    \includegraphics[width=0.49\textwidth,trim={15cm 5cm 15cm 4.1cm},clip]{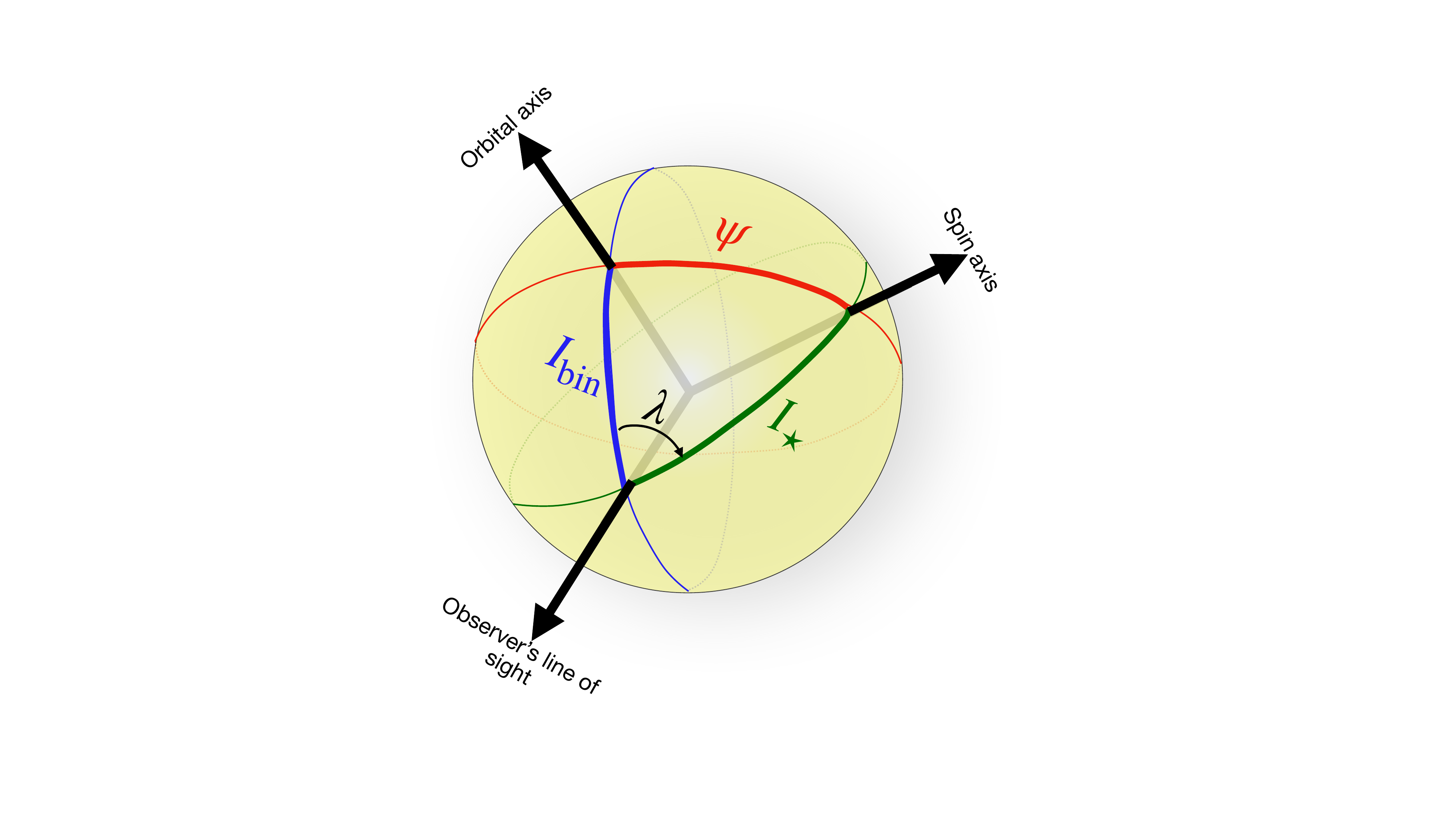}
    \caption{Orbital geometry of stellar obliquities adapted from a figure in \citet{Albrecht2021}. The diagram shows $\psi$ - the true stellar obliquity, $\lambda$ - the sky-projected spin-orbit angle, $I_{\rm \star}$ - the inclination of the rotation axis of the primary star, and $I_{\rm bin}$ - the orbital inclination of the binary.}\label{fig: Orbital Geo}
\end{figure}

\subsubsection{Secondary Star Effective Temperature}

The depth of the secondary eclipse (Fig.~\ref{fig: Eclipses and Light curves}) is indicative of the surface brightness ratio ($S_{\rm B}/S_{\rm A}$) between the two stars. This is not the same as the effective temperature ratio. Since we are only observing the eclipses in one photometric bandpass, and hence have no colour information, we have to be creative to estimate $T_{\rm eff, B}$. We follow the same procedure as done previously by \citet{Swayne2020,Swayne2021,Canas2022, Duck2023, Spejcher2025b}. The \textit{xo.SecondaryEclipseLightCurve} function in our \textsc{exoplanet}/\textsc{starry} MCMC directly constrains the stellar radius ratio ($k=R_{\rm B}/R_{\rm A}$) and the secondary eclipse depth ($D_{\rm sec}$). These are related to $T_{\rm eff, B}$ via

\begin{align}
    \label{eq:sec_eclipse_depth}
    D_{\rm sec} &=k^2S + A_{\rm g} \left(\frac{R_{\rm B}}{a}\right)^2, \\
    \label{eq:sec_eclipse_depth2}
    &= \left(\frac{R_{\rm B}}{R_{\rm A}}\right)^2\frac{\int \tau(\lambda)F_{\rm B,\nu}(\lambda,T_{\rm eff,B})\lambda d\lambda}{\int \tau(\lambda)F_{\rm A,\nu}(\lambda,T_{\rm eff,A})\lambda d\lambda} + A_{\rm g} \left(\frac{R_{\rm B}}{a}\right)^2,
\end{align}
where $A_{\rm g}$ is the geometric albedo, $\tau(\lambda)$\footnote{\url{http://svo2.cab.inta-csic.es/svo/theory/fps3/index.php?id=Kepler/Kepler.K}} is the TESS transmission function as a function of wavelength $\lambda$ and $F$ is the normalised flux. For eclipsing M-dwarfs, $A_{\rm g}$ is close enough to zero and the secondary star temperature is hot enough such that the second term of Eq.~\ref{eq:sec_eclipse_depth2} can be ignored.

We calculate the integrated flux of the primary star using a high-resolution PHOENIX model spectrum, convolved with the TESS bandpass. The model spectrum is based on the  best-fit values for $T_{\rm eff,A}$, $\log g_{\rm A}$ and [Fe/H] from \textsc{EXOFASTv2}. We then calculate the brightness of the secondary star across a grid of potential effective temperatures between 2300 and 4000 K. Using Brent minimisation we solve for $T_{\rm eff,B}$. This process is repeated with values of the radius ratio ($R_{\rm B}/R_{\rm A}$) and secondary eclipse depth ($D_{\rm sec}$) randomly drawn from $1\sigma$ errors of the MCMC-fitted parameters. From this we derive a posterior and errorbar on $T_{\rm eff,B}$.

% \begin{figure*}
%     \includegraphics[width=0.99\textwidth]{RV Params Corner Plot.pdf}
%     \caption{Corner plot produced from the MCMC simulation of the joint fit to all RVs (Keplerian and Rossiter) and the TESS photometry for primary and secondary eclipses. The plot shows the correlation between the parameters corresponding to the RM fit - $v_{\rm rot} \sin I_{\rm star}$, $\lambda$, $\psi$, \& $P_{\rm rot}$.}\label{fig: Corner Plot 4 Params}
% \end{figure*}

\subsection{Fitting the Rossiter-McLaughlin Anomaly}

We use the RM effect to measure the alignment between the stellar spin axis and the orbital axis in our systems. Since EBLM has a very uneven flux ratio ($F_{\rm B}/F_{\rm A}\sim0.0015$), we are inherently referring to the spin-axis of the primary star. At its core, The RM effect is a distortion of stellar spectral lines during a transit or eclipse, caused by the foreground object (in our case, an M-dwarf) blocking rotating regions of the background star (F/G/K dwarf) and creating an asymmetric Doppler profile. Traditional methods for RM fitting constrain the projected rotational velocity of the occulted star, $v_{\rm rot} \sin I_{\rm \star}$, and the magnitude of the projected spin-orbit obliquity, $|\lambda|$. \citet{Hirano2011} updated RM models to account for macroturbulence ($\zeta$), thermal broadening ($\beta$) and Lorentzian broadening (natural and pressure ($\gamma$). We combine the \citet{Gupta:2024} publicly-available implementation of the \citet{Hirano2011} model into our \textsc{exoplanet} joint fit of the Keplerian RVs and photometric eclipses. This is similar to what \citet{Carmichael2025} used to fit the RM effect of brown dwarfs with {\tt \textsc{EXOFASTv2}} and is the same technique used in \citet{Spejcher2025b}. In Fig. \ref{fig:RM Example Graph} we show the fitted RM anomaly's of our systems.

The code from \citet{Gupta:2024} outputs the \textit{sky-projected obliquity ($\lambda$)}. This tells us how misaligned our system is as seen from Earth. For any system that has an observed RM anomaly, we can fit for $\lambda$. For most systems, this is the only obliquity that can be measured. However, if a star has a known rotational period, we can calculate the \textit{true obliquity ($\psi$)} using the equation

\begin{equation} \label{True Obl}
    \cos\psi \approx \cos\lambda \frac{P_{\rm rot}v_{\rm rot}\sin I_{\rm \star}}{2\pi R_{\rm A}},
\end{equation}
as is the case with our five EBLM targets. As seen in Fig. \ref{fig: Orbital Geo}, $\lambda$ is the angle between the sky projections of the orbital and rotational axes, where $\psi$ is measured between the axes in three dimensions \citep{Albrecht2021}.

The stellar rotation period of each system's primary star was measured by \citet{Sethi2024} using spot modulations. Since star spots are cooler and dimmer than the surrounding photosphere, the presence of a spot overdensity leads to periodic variations in brightness as the star rotates. By measuring the period of the out-of-eclipse variability in the TESS light curves, \citet{Sethi2024} was able to determine the rotational periods of our systems, given in Table \ref{table:MCMC Params}. We include this in our model as a Gaussian and use this to solve for the inclination. This will allow us to measure the true obliquity of the system using Eq. \ref{True Obl}. 

%Using this value as a prior, we fit for the rotation period using a normal distribution. 

We use the Angle function from \textsc{Pymc3 Extras} to find the best $\lambda$ value within the range of $-\pi$ to $\pi$. This value is then fed into the Keplerian orbit model to fit the anomaly. We then use our fitted value for the rotation period to calculate the equatorial rotation velocity, where $v_{\rm rot} =2\pi R_{\rm A}/P_{\rm rot}$. With this we can find $\sin I_{\rm \star}$, where $I_{\rm \star}$ is the stellar inclination, using the equation:

\begin{equation}
    \sin I_{\rm \star} = \frac{v_{\rm rot} \sin I_{\rm \star}}{v_{\rm rot}}.
\end{equation}

With these parameters set, we then use Eq. \ref{True Obl} to calculate $\psi$ as a deterministic variable. The final values for $\psi$ and $\lambda$ along with all the other parameters from the RM, RV, and photometry joint fit can be found in Table \ref{table:MCMC Params}. The distributions of $\lambda$, $\psi$, $P_{\rm rot}$, and $v_{\rm rot} \sin I_{\rm \star}$ for each system can be found in Fig. \ref{fig: Corner Plot 1} - \ref{fig: Corner Plot 5}.

\section{Results}\label{section:results}

Our best-fitted parameters for each system are shown in Table \ref{table:MCMC Params}. Additionally, a comparison of our results to previous EBLM systems is shown in Table \ref{EBLM Target Comp}. Using the methods outlined in Sect. \ref{section:methods} for modeling the RM anomaly, we found EBLM J0239-20 to have a sky-projected obliquity of $|\lambda| = 4.2 \pm 2.2^{\circ}$, and a true 3D obliquity of $\psi = 7.9 \pm 1.2^{\circ}$; EBLM J0941-31 to have a sky-projected obliquity of $|\lambda| = 11.7 \pm 4.8^{\circ}$, and a true 3D obliquity of $\psi = 13.6 \pm 5.4^{\circ}$; EBLM J1037-25 to have a sky-projected obliquity of $|\lambda| = 2.64 \pm 1.3^{\circ}$, and a true 3D obliquity of $\psi = 16.5 \pm 5.5^{\circ}$; EBLM J1141-37 to have a sky-projected obliquity of $|\lambda| = 0.32 \pm 0.69^{\circ}$, and a true 3D obliquity of $\psi = 18.0 \pm 4.7^{\circ}$; and EBLM J2025-45 to have a sky-projected obliquity of $|\lambda| = 0.7 \pm 1.7^{\circ}$, and a true 3D obliquity of $\psi = 17.3 \pm 6.8^{\circ}$. EBLM J0239-20 and EBLM J1141-37 were measured to have nearly circular orbits ($e<0.005$) while EBLM J0941-31, EBLM J1037-25, and EBLM J2025-45 are on more eccentric orbits ($0.1<e<0.2$). Our fits to the RM effect can be seen in Fig. \ref{fig:RM Example Graph} with a graphic showing the trajectory of the M dwarf around the primary star based on our values of $\psi$.

\section{Discussion} \label{section:discussion}

\subsection{Tidal Effects}\label{subsec:tides}

\begin{table} 
\caption{Orbital period ($P_{\rm orb}$), secondary mass ($M_{B}$), projected obliquity $|\lambda|$, and true obliquity ($\psi$, if measured) for all EBLM targets with measured RMs. This includes the brown dwarf WASP-30, which is also known as EBLM J2353-10.}              % title of Table
\label{EBLM Target Comp}      % is used to refer this table in the text
\centering                                    % used for centering table
%\resizebox{0.95\linewidth}{!}{
\begin{tabular}{m{1.15cm} p{1.4cm} p{0.9cm} p{0.8cm} p{0.5cm} p{1.4cm}}
% \begin{tabular}{c c l l l l l}  

\hline\hline                        % inserts double horizontal lines
 
 Target & $P_{\rm orb}$ & $M_{B}$ & $|\lambda|$ & $\psi$ & Citation\\

  & [days] & [$M_{\odot}$] & [$^\circ$] & [$^\circ$] & \\
 
\hline 

WASP-30 & 4.156739 $\pm0.000011$ & 0.0597 $\pm0.0011$ & 7 $\pm23$ & & \citet{Triaud2013} \\
EBLM J1219-39 & 6.7600098 $\pm0.000028$ & 0.0910 $\pm0.0021$ & 4.1 $\pm5$ & & \citet{Triaud2013} \\
EBLM J0218-31 & 8.884102 $\pm0.000011$ & 0.390 $\pm0.009$ & 4 $\pm7$ & & \citet{Gill2019} \\
EBLM J0608-59 & 14.608559 $\pm0.000013$ & 0.313 $\pm0.011$ & 2.8 $\pm17.1$ & & \citet{hodzic2020} \\
EBLM J0021-16 & 5.96727088 $\pm0.00000034$ & 0.1877 $\pm0.0011$ & 2.0 $\pm1.1$ & 28.9 $\pm 2.1$ & \citet{Spejcher2025b} \\
EBLM J0239-20 &  2.77868784 $\pm0.00000034$ &  0.17416 $\pm 0.00092$ &  4.2 $\pm2.2$ & 7.9 $\pm 1.2$ & Spejcher et al. (this work) \\
EBLM J0941-31 & 5.54564703 $\pm0.00000045$ &  0.2309 $\pm 0.0057$ &  11.7 $\pm 4.8$ &  13.6 $\pm5.4$ & Spejcher et al. (this work) \\
EBLM J1037-25 & 4.93655905 $\pm0.00000021$ &  0.2755 $\pm 0.0013$ &  2.64 $\pm 1.3$ &  16.5 $\pm 5.5$ & Spejcher et al. (this work) \\
EBLM J1141-37 & 5.147689047 $\pm0.000000099$ & 0.3679 $\pm 0.0066$ &  0.32 $\pm0.69$ &  18.0 $\pm 4.7$  & Spejcher et al. (this work) \\
EBLM J2025-45 & 6.19198051 $\pm0.00000025$ &  0.2239 $\pm 0.0014$ &  0.7 $\pm 1.7$ & 17.3 $\pm 6.8$ & Spejcher et al. (this work) \\

\hline

\end{tabular}

\vspace{1mm}
{\raggedright 
\small \textit{Note ---} \citet{Triaud2013} accounted for asymmetrical errors in their fits of WASP-30 and EBLM J1219-39, so the errors given here are the average of the asymmetric errors given in \citet{Triaud2013}. This was done to provide a rough symmetry of errors to match the errors from our calculation, which assumes a $1\sigma$ symmetry.\par}

\end{table}

For binary systems with sufficient angular momentum, tides act to bring the system to three equilibrium states \citep{Hut1980, O2014, Barker2025}:

\begin{enumerate}
    \item spin-orbit alignment: the true obliquity $\psi=0^{\circ}$;
    \item spin-orbit synchronization: the primary and secondary stellar rotation rate equals the orbital period, $P_{\rm rot}=P_{\rm orb}$;
    \item circularization: the orbital eccentricity $e=0$.
\end{enumerate}

In this paper, we measured five EBLM systems using a joint MCMC fit to measure the eccentricity and true obliquity of the systems. The true obliquity for each system shows slight spin-orbit misalignment, with $\psi \approx 15^{\circ}$. EBLM J0239-20 is an exception with $\psi=7.9\pm1.2^{\circ}$. Note that we can only speak of the spin of the primary star for each system. 

\citet{Sethi2026} calculated the spin-orbit synchronization and circularization timescales for 68 EBLM systems, including the five systems we look at here. These values can be found in the last rows of Table \ref{table:MCMC Params}.

The synchronization timescale was calculated using the following equation from \citet{Barker2020}:

\begin{equation} \label{Synchronization Eq}
    \tau_{\rm sync} = \frac{2Q'r_{\rm g}^2}{9\pi}\left(\frac{M_{\rm A}+M_{\rm B}}{M_{\rm B}}\right)^2 \frac{P_{\rm orb}^4}{P_{\rm dyn}^2P_{\rm rot}},
\end{equation}
where $P_{\rm dyn} = 2\pi\sqrt{R_{\rm A}^3/(GM_{\rm A})}$ is the dynamical timescale of the primary star and $r_{\rm g}^2$ is its squared radius of gyration (see \citealt{Barker2020} for further details). Here, $Q'$ represents the dissipation of inertial waves according to the frequency-averaged formalism of \citet{Ogilvie2013}. This accounts for the realistic structure of the star following \citet{Barker2020, B2022}. For solar-type main-sequence stars with masses less than $1.1 M_{\odot}$, a typical value for $Q'$ is $Q'\sim 10^7(P_{\rm rot}/10 \rm d)^2$ and is the simplest way to represent inertial wave dissipation (see \citealt{LO2017}, \citealt{Barker2020}, and \citealt{Sethi2026}).  

Spin-orbit synchronization occurs when $P_{\rm orb}/P_{\rm rot} = 1$. Within $2\sigma$ of the rotational period, EBLM J0239-20 (Age$=6.4\pm2.5$ Gyr), EBLM J1037-25 (Age$=6.12\pm1.02$ Gyr), and EBLM J1141-37 (Age$=2.63\pm 1.52$ Gyr) have synchronous orbits. This is in alignment with the calculated synchronization timescales from \citet{Sethi2026}, who found the timescales for these systems to be between $10^5$ and $10^7$ years. Although the errors on the synchronization timescales for EBLM J0239-20 and EBLM J1037-25 are greater than two orders of magnitude of their calculated timescales, the timescales are still shorter than the age of the systems ($\tau_{\rm sync} \sim 10^5$yr and $\tau_{\rm sync} \sim 10^6$yr, respectively). EBLM J1141-37 has synchronization timescale uncertainty that is less that two orders of magnitude of its calculated timescale ($\tau_{\rm sync} \sim 10^7$), meaning that EBLM J1141-37 \textit{should be} spin-orbit synchronized, which is consistent with our results. EBLM J0941-31 (Age$=2.4\pm 0.77$ Gyr) is a supersynchronous system, with $P_{\rm orb}/P_{\rm rot} > 1$, and EBLM J2025-45 (Age$=10.4\pm 2.9$ Gyr) is a subsynchronous system with $P_{\rm orb}/P_{\rm rot} < 1$. The timescales of synchronization for EBLM J0941-31 and EBLM J2025-45 are on the order of $10^7$ and $10^6$ years, respectively, so we expect these systems to be synchronized. However, the error on the synchronization timescale for EBLM J0941-31 is larger than 2 orders of magnitude, meaning that within the uncertainty, the timescale for spin-orbit synchronization may be longer than the current age of the system \citep{Sethi2026}

It is worth noting that the classical pseudo-synchronous rotation period for eccentric orbits with small $e$ can be approximated by 

\begin{equation}
P_{\rm rot, ps} \approx \frac{P_{\rm orb}}{1+6e^2}.
\end{equation}
For pseudo-synchronisation to occur, the rotation speed of the primary star must be approximately equal to the orbital speed at periapsis. For our three eccentric systems, we find that $P_{\rm rot, ps} = 4.49$ days for EBLM J0941-31, suggesting that the secondary is slowing down as it orbits the primary, which is consistent with pseudo-synchronization and may be a way to explain the supersynchronous orbit that we measure. EBLM J1137-25 has a pseudo-synchronous rotational period of $P_{\rm rot, ps} = 4.54$ dyas, but has observed rotational and orbital periods that are consistent with full synchronization. And for EBLM J2025-45, we find a pseudo-synchronous rotational period of $P_{\rm rot, ps} = 5.65$ days compared to an observed rotational period of $P_{\rm rot} = 6.43$ days and an orbital period of $P_{\rm orb} = 6.19$ days. This system is rotating slower than expected at periapsis. We are hesitant to draw strong conclusions from a single system though, and note that differential rotation could confuse our observed rotation periods.

To calculate the timescales for spin-orbit alignment, we use the equation from \citet{Lai2012}, while following the methods from \citet{O2014, B2016, Barker2020}, to find the realignment time scale:
\begin{equation} \label{Realignment Eq}
    \tau_{\psi} = \frac{4Q'r_{\rm g}^2}{9\pi}\left(\frac{M_{\rm A}+M_{\rm B}}{M_{\rm B}}\right)^2 \frac{P_{\rm orb}^4}{P_{\rm dyn}^2P_{\rm rot}}=2 \tau_{\rm sync},
\end{equation}
assuming a similar $Q'$ applies to damp obliquity tides as $l=m=2$ tides. This is approximately the case by comparing \citet{LO2017} with the corresponding $l=m=2$ case.

The computed timescales for realignment for our sample range from $10^5$ to $10^7$ years, which is smaller than the MIST stellar ages of all of the systems, which are on the order of $10^9$ years. This suggests that all of our systems should be spin-orbit aligned, yet our results show slight misalignment for all of our systems.

The circularization timescales were calculated for both circularization driven by primary dissipation and circularization driven by secondary dissipation using the equation given by \citet{Barker2020} \citep{Sethi2026}:
\begin{equation} \label{Circularization Eq}
    \tau_{e, \rm A} = \frac{2Q'}{63\pi}\left(\frac{M_{\rm A}}{M_{\rm B}}\right)\left(\frac{M_{\rm A} + M_{\rm B}}{M_{\rm A}}\right)^{\frac{5}{3}}\frac{P_{\rm orb}^{\frac{13}{3}}}{P_{\rm dyn}^{\frac{10}{3}}},
\end{equation}
where $P_{\rm dyn}$ for the primary star is the same as defined previously. To calculate $\tau_{e, \rm B}$, we can interchange subscripts A and B in Eq. \ref{Circularization Eq} and compute $P_{\rm dyn}$ for the secondary star.

For each of our systems, the timescales for circularization driven by primary dissipation were found to be more efficient than secondary dissipation ($\tau_{e, \rm A}\sim10^5-10^8$ years versus $\tau_{e, \rm B}\sim10^8-10^9$ years). Additionally, all systems are expected to have circularized orbits if we take the minimum circularization timescale ($\tau_{e, \rm A}$ in our case). Our sample contains two systems with nearly circular orbits (EBLM J0239-20, EBLM J1141-37), which is expected when looking at both primary and secondary timescales from \citet{Sethi2026}. EBLM J1037-25 has circularization timescales of $\tau_{e, \rm A}=3.52\times10^7$ years and $\tau_{e, \rm B}=8.28\times10^8$ years, so we expect a circularized orbit (recall that the age of the system is $6.5\pm1.02 \times10^9$ years). However, we measure the eccentricity of the system to be modest, with $e=0.12139\pm0.000075$. Furthermore, EBLM J0941-31 and EBLM J2025-45 have secondary star driven circularization timescales that are greater than the age of the system, or very close to the age of the system. This is more consistent with the eccentricity that we are finding where EBLM J0941-31 has an eccentricity of $e=0.19853 \pm 0.00014$ and EBLM J2025-45 has an eccentricity of $e=0.12651 \pm 0.00016$. However, it is important to note that the timescales of primary circularization for EBLM J0239-20, EBLM J0941-31, and EBLM J1037-25 have uncertainties in the timescales larger than 2 orders of magnitude from the calculated timescale. These large uncertainties come from the large uncertainties in the ages of the system \citep{Sethi2026}. EBLM J1141-37 and EBLM J2025-45 have uncertainties that are less than 2 orders of magnitude of the calculated timescales, so less caution is needed in interpreting these results.

\subsection{Obliquity Across the Spectrum of Mass Ratio} 

\begin{figure*}
    \includegraphics[width=0.49\textwidth]{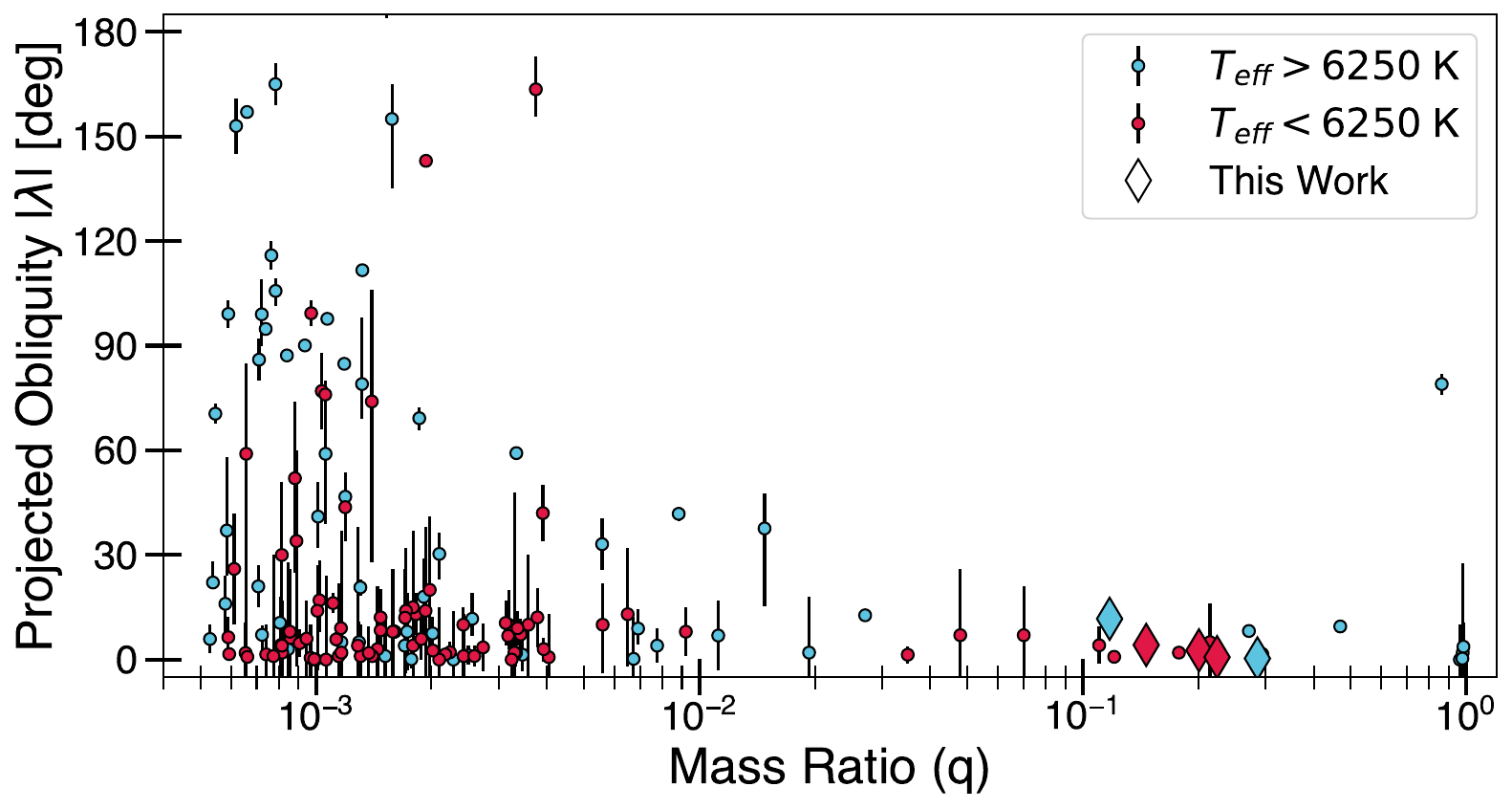}
    \includegraphics[width=0.49\textwidth]{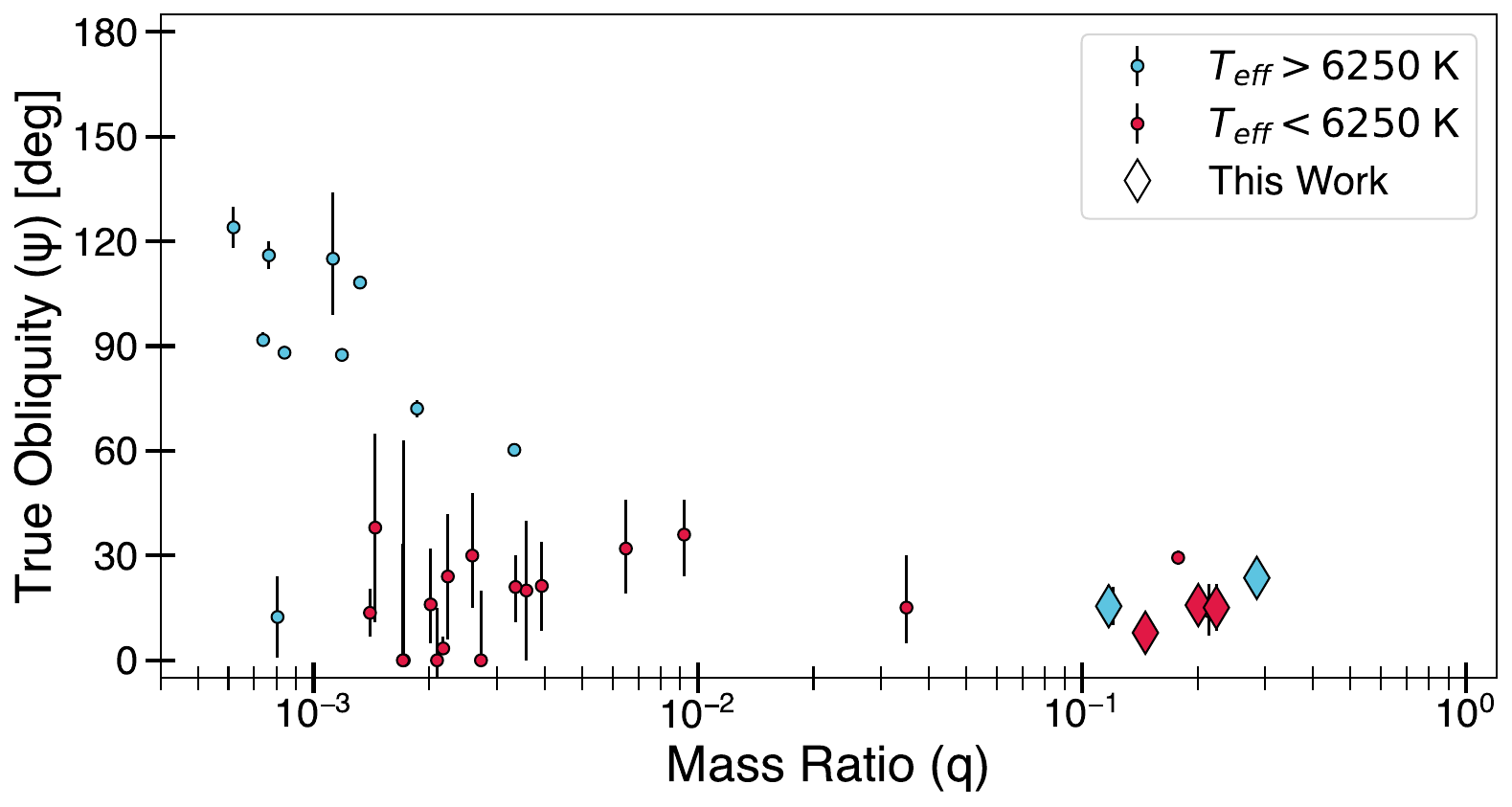}
    \caption{Sky-projected stellar obliquity $\rm |\lambda|$ vs. mass ratio (q, left) and true 3-D Stellar Obliquities ($\psi$) vs mass ratio (right) for eclipsing companions, $M_b > 0.7$ $M_{\rm J}$. Obliquities measured via apsidal motion are not depicted. Red points depict systems with primary stars below the Kraft break \citep[$\sim6250$ $K$;][]{Kraft1967}, and blue points depict those above. The new systems we add are shown as diamonds. The giant planet and brown dwarf obliquities were retrieved from TEPCat \citep{Southworth2011} and \citet{Carmichael2025, Vowell2025b}. The eclipsing binary obliquities are from \citet{Marcussen2022, Spejcher2025b, Wells2025}.}
    \label{fig:lambda-q}.
\end{figure*}

Table \ref{EBLM Target Comp} shows the orbital period ($P_{\rm orb}$), secondary mass ($M_{\rm B}$), sky-projected obliquity $|\lambda|$, and true obliquity ($\psi$) for each EBLM system with a measured RM, including the brown dwarf WASP-30 and the 5 systems discussed in this work. Note that the only other stellar binary with a measured true obliquity before this work is EBLM J0021-16 from \citet{Spejcher2025b}. 

In Fig. \ref{fig:lambda-q}, we show the projected and true obliquities for all giant planets, brown dwarfs, and stellar companions with $M_{\rm b} > 0.7 M_{\rm J}$ \citep{Southworth2011,Marcussen2022,Carmichael2025,Vowell2025b}. Smaller mass ratios, concentrated around $q=10^{-3}$, correspond to RM measurements of hot Jupiters. We can see that roughly 1/3 of the hot Jupiter population is misaligned \citep{Triaud2010, Winn2015}. Note that a majority of the misaligned hot Jupiters have primary star temperatures greater than the Kraft break \citep[$\sim6250$ $K$;][]{Kraft1967}. This is likely because stars above the Kraft break have radiative outer envelopes, which is thought to be less efficient for tidal realignment. Hot Jupiters with primary stars below the Kraft break tend to be more aligned, which corresponds to stars with thick convective outer envelopes and more efficient tidal effects.
Another trend that emerges when looking at the obliquity of systems with different mass ratios is that systems with larger mass ratios tend to be more aligned \citep{Rusznak2025, Carmichael2025, Vowell2025b}. This is likely due to tides being more efficient in systems with higher mass ratios. This can be seen in Eq. \ref{Realignment Eq}, which shows that the realignment timescale for stars with convective envelopes scale as $\tau_{\rm align, CE}\propto1/q^2$, which suggests eclipsing binaries (high $q$) might all be aligned. \citet{Zahn1977} also shows that there is an even stronger relationship between alignment timescales and mass ratio for hotter stars, with radiative outer envelopes. For systems with host stars above 1.6$M_{\odot}$, the realignment timescale scales as $\tau_{\rm align, RE}\propto1/(q^2(1+q)^{5/6})$.

It is important to note that we define the Kraft break to be $T_{\rm eff}=6250K$, which is consistent with traditional obliquity studies. A recent paper by \citet{Wang2025} shows that if you consider obliquity measurements for single star systems only, the obliquity transition between aligned and misaligned systems moves closer to $\sim6500K$, which is more consistent with the rotational Kraft break. However, if you consider single star \textit{and} multiple star systems with obliquity measurements, you find the obliquity transitions sits at the traditional $6250K$ value that we are using. In our sample, EBLM J0941-31A and EBLM J1141-37A both have effective temperatures greater than the traditional Kraft break ($T_{\rm A} = 6450K$ and $T_{\rm A} = 6390K$, respectively). The measured true 3D obliquities for these systems are both slightly misaligned. Our other 3 systems have primary effective temperatures under the Kraft break. Additionally, when \citet{Wang2025} describes single star systems, they are referring to just one object transiting a primary star. A multiple star system describes an object transiting a primary star with a long-period secondary stellar companion that is also orbiting the primary star. For our sample, a single star system is equivalent to our short-period eclipsing binaries with no known tertiary body compared to our short-period eclipsing binaries \textit{with} a tertiary body. EBLM J0239-20 and EBLM J0941-31 both have a linear trend, indicative of a tertiary body. EBLM J0239-20 has a primary effective temperature below the Kraft break and is our most aligned system in our sample. EBLM J0941-31 has a primary effective temperature above the Kraft break and is expected to be misaligned in the context of \citet{Wang2025}, however we only see slight misalignment.

We encourage caution with respect to interpreting these early measurements  of tight stellar binary obliquities. Understanding how these binaries form and evolve will be an interesting challenge for more complex tidal models to describe the misalignment trends that we see here. More RM measurements, especially for more systems above the Kraft break and for longer period binaries, where tidal effects are slower, will shed more light on misalignment timescales.

\subsection{Mass-Radius Relationship}

Another of the EBLM survey is the precise characterization of M-dwarfs in eclipsing binaries \citep{Maxted2023}. The goal is to measure their mass and radius to a precision of better than a couple of per cent. More precisely-measured M-dwarf parameters are crucial to our understanding of stellar structure and evolution models. Furthermore, the reliability of our planetary parameters is dependent on the reliability of the stellar parameters. This requires better M-dwarf models, both theoretical and empirical. 

We measure the mass and radius of our systems to less-than 3\% uncertainties. It is important to note that these are nominal uncertainties and that the true uncertainties could be larger due to unaccounted for systematics. One source of systematic uncertainty comes from our derivation of primary star parameters. Since each system is a single-lined spectroscopic binary, we do not get model independent masses. However, \citet{Duck2023} showed for two EBLM systems that systematic model uncertainty can be similar to the nominal uncertainty.

\begin{figure*}
    \includegraphics[width=0.995\textwidth,trim={3cm 0.1cm 3cm 0.1cm},clip]{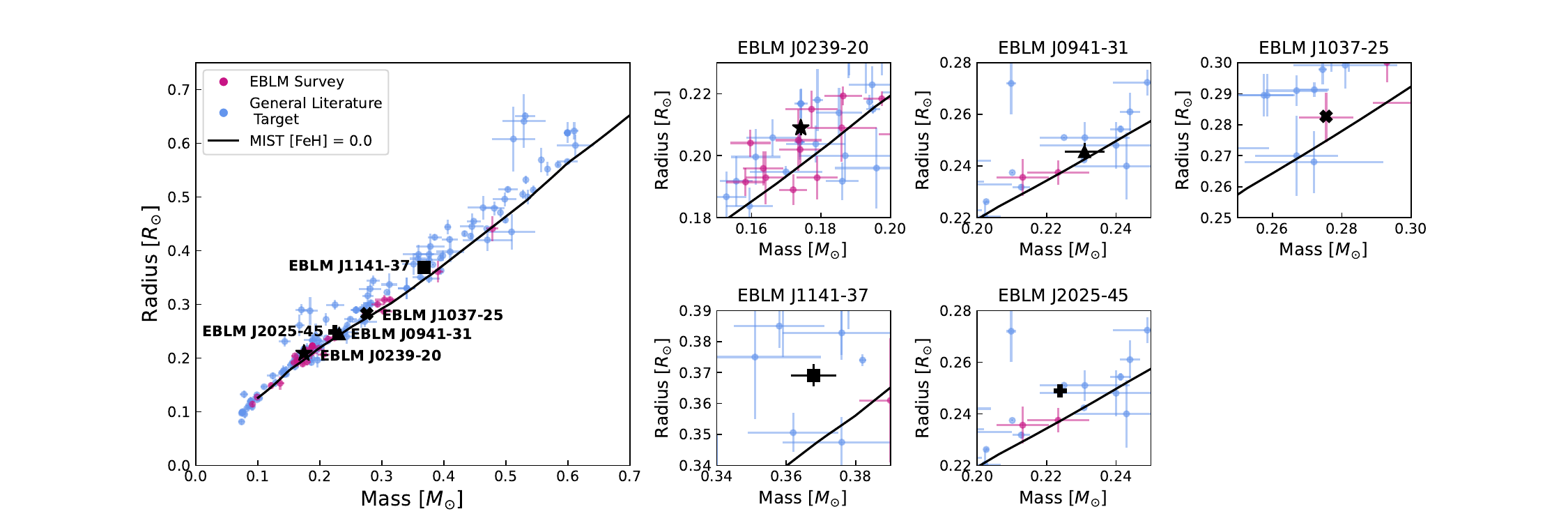}
    \caption{Mass vs. Radius plot for high-precision M-dwarfs, with data taken from \citet{Maxted2023}. The plot includes targets from the EBLM survey (pink dots) and general literature targets (blue dots). Only targets with a mass or radius better than 10\% are included here. The solid line passing through the data points indicates the MIST stellar model for [Fe/H] = 0.0. Our systems are highlighted by the black markers.} \label{fig: Mass vs Radius}
\end{figure*}

Fig. \ref{fig: Mass vs Radius} shows the distribution of mass and radius of the well-characterized M-dwarf population. The mass-radius relationship obtained from the MIST stellar model with [Fe/H] = 0.0 is also shown. Among this population, many M-dwarfs are observed to have larger radii than what was expected from the model. This was dubbed the ``radius inflation'' problem \citet{Parsons2018, Morrell2019}. The secondary stars in our sample are shown by the black markers, each of which lie above the theoretical prediction. To quantify the radius inflation, we use the following equation from \citet{Gill2019}:
\begin{equation} \label{Radius Inflation}
    \frac{\Delta R_{\rm B}}{R_{\rm B}} = \frac{R_{\rm B} - R_{\rm B, exp}}{R_{\rm B}},
\end{equation}
where $R_{\rm B}$ is the radius of the M-dwarf given by the MCMC fit and $R_{\rm B, exp}$ is the expected radius value from the MIST model. To find the expected radii of our systems, we interpolate over the MIST model using the mass of the M-dwarfs given by the MCMC joint fit (see Table \ref{table:MCMC Params}). Fig. \ref{fig: Mass vs Radius} (right) shows where each of our systems lie with respect to the MIST stellar model. It is clear that each system except for EBLM J0941-31B lies well above the model. These systems are all inflated by greater than $5\sigma$ from their expected radius values. EBLM J0941-31B is not inflated, as it is within $1\sigma$ from the expected radius value based on the MIST stellar model. 

\section{CONCLUSION}\label{section:conclusion}
We combined \coralie ~and HARPS spectroscopic data with TESS photometry to characterize a sample of five EBLM systems. Within our \coralie ~RVs contain observations during primary eclipses for the systems. Additionally, TESS photometry shows strong star spot modulation, making it possible to measure the rotation period of the primary star in each system \citep{Sethi2024}. This combination makes it possible to precisely characterize the Rossiter-McLaughlin effect in these systems, both the sky-projected and true 3D obliquity. We reveal that whilst the sky-projected obliquity for three systems suggests spin-orbit alignment ($\lambda<3^{\circ}$), the true 3D obliquity tends to be slightly misaligned ($\psi>15^{\circ}$). For EBLM J0239-20, we measure a sky-projected obliquity of $\lambda=4.2\pm2.2^{\circ}$ and 3D obliquity of  $\psi=7.9\pm1.2^{\circ}$. For EBLM J0941-31, we measure a sky-projected obliquity of $\lambda=11.7\pm4.8^{\circ}$ and a 3D obliquity of $\psi=15.5\pm6.1^{\circ}$. Both of these systems have 3D obliquities that are more consistent with their sky-projected obliquities. It is interesting to note that both of these systems have a possible tertiary companion based on a necessary linear trend that needed to be added to their joint RV and photometry fits. Additionally, EBLM J0239-20 has a primary effective temperature below the Kraft break ($\sim 6250K$), has a nearly circular orbit ($e=0.00198\pm0.00056$), and tends to be more aligned than EBLM J0941-31, which has a primary effective temperature above the Kraft break and a more elliptical orbit ($e=0.19852\pm0.00013$).

Prior to this work, only five EBLM systems have measured sky-projected obliquities, and only one with a measured 3D obliquity \citep{Spejcher2025b}. We have now doubled the number of EBLM systems with measured RM anomalies and quintupled the number of systems with measured true 3D obliquities. In total, there are now 13 main sequence eclipsing binaries with RM observations and measured sky-projected obliquities that show a majority are (or are close to being) spin-orbit aligned. This is in contrast to the 132 giant planets that have RM measurements and show a diversity of obliquities. Additionally, brown dwarf obliquity measurements show a similar alignment trend as short period stellar binaries. This may be due to tidal realignment in hot Jupiters being less efficient in systems with high mass ratios. Alternatively, we find that although the sky-projected obliquities show near alignment for most of our systems, the true 3D obliquities show a slight misalignment. Our sample also provides two systems with primary effective temperatures above the Kraft break, both of which are slightly misaligned. This is expected to be a consequence of tidal realignment being less effective in stars with radiative outer envelopes. It is possible that more massive companions like brown dwarfs and M-dwarfs tend to be more aligned because they migrate differently than hot Jupiters (e.g. disk migration vs. high eccentricity todal migration). To distinguish between whether more efficient tides or a difference in migration causes alignment for systems with higher mass ratios, we encourage further RM observations, particularly for eclipsing binaries with longer periods and hotter primary stars, where tides are expected to be weaker.

Finally, we measure the mass and radius of the M-dwarfs in our sample to a precision better than 3\%. Comparing these values to the MIST stellar model with $[\rm Fe/ H]=0.0$, we find that 4 out of 5 M-dwarfs have radii inflated by greater than $5\sigma$. The exception to this is EBLM J0941-31, which has a radius that lies within $1\sigma$ from the expected value. This precise M-dwarf radius measurements add five more pieces to the radius inflation puzzle.

\section*{Acknowledgements}

The EBLM Program has been supported by NASA TESS Guest Investigator Program through Cycle 1 (G011278), Cycle 2 (G022253), Cycle 3 (G03195), Cycle 4(G04157), Cycle 5 (G05073), Cycle 6 (G06022), Cycle 7 (G07005) and Cycle 8. Postage stamp observations were provided in all 7 cycles and funding was provided in cycles 2, 4, 5 and 7. This research is also supported work funded from the European Research Council (ERC) the European Union’s Horizon 2020 research and innovation programme (grant agreement n◦803193/BEBOP). AJB was supported by STFC grants ST/W000873/1 and UKRI1179. 

This paper is the result of work stemming from the AST-192 Research For Credit class in Spring 2025.

The authors would like to attract attention on the help and kind attention of the ESO staff at La Silla and on the dedication of the many technicians and observers from the University of Geneva, to upkeep the telescope and acquire the data that we present here. We would also like to acknowledge that the Euler Swiss Telescope at La Silla is a project funded by the Swiss National Science Foundation (SNSF). 

\section*{Data Availability Statement}

All radial velocities and light curves will be made available online.

\renewcommand{\arraystretch}{1.5}
\begin{table*}
\caption{Final fitted parameter values for the joint Keplerian, RM, and photometric fit. We also include the timescales for synchronization and circularization as calculated in Sethi et al. (in prep) using Eq. \ref{Synchronization Eq} and \ref{Circularization Eq} respectively. The realignment timescale was calculated using Eq. \ref{Realignment Eq} from \citet{B2016}. The parameters are separated by physical, observable, and RM-specific parameters.} \label{table:MCMC Params}
\centering                                  
\begin{tabular}
{m{1.5cm}>{\raggedright\arraybackslash}m{2.6cm}m{2.5cm}m{2.5cm}m{2.5cm}m{2.5cm}m{2.5cm}}

\hline
Parameter & Description & EBLM J0239-20 & EBLM J0941-31 & EBLM J1037-25 & EBLM J1141-37 & EBLM J2025-45\\
\hline \hline

\multicolumn{7}{c}{Physical Parameters} \\ \hline
$m_{\rm A}$ ($M_{\odot}$) & Primary Mass & $1.1908_{- 0.0099}^{+0.0098}$ & $1.1672_{- 0.056}^{+0.059}$ & $1.3721_{- 0.01}^{+0.01}$ & $1.2929_{- 0.038}^{+0.039}$ & $1.0011_{- 0.01}^{+0.01}$ \\
$m_{\rm B}$ ($M_{\odot}$) & Secondary Mass & $0.17416_{- 0.00093}^{+0.00092}$ & $0.21547_{- 0.0064}^{+0.0068}$ & $0.2755_{- 0.0013}^{+0.0013}$ & $0.36836_{- 0.0066}^{+0.0068}$ & $0.22399_{- 0.0014}^{+0.0014}$ \\
$q$ & Mass Ratio & $0.14626_{- 0.00047}^{+0.00045}$ & $0.18478_{- 0.0035}^{+0.0033}$ & $0.20079_{- 0.00055}^{+0.00055}$ & $0.28502_{- 0.0034}^{+0.0033}$ & $0.22375_{- 0.00087}^{+0.00084}$ \\
$R_{\rm A}$ ($R_{\odot}$) & Primary Radius & $1.635_{- 0.014}^{+0.014}$ & $1.766_{- 0.027}^{+0.029}$ & $1.722_{- 0.0077}^{+0.0071}$ & $1.806_{- 0.016}^{+0.017}$ & $1.053_{- 0.0053}^{+0.005}$ \\
$R_{\rm B}$ ($R_{\odot}$) & Secondary Radius & $0.2092_{- 0.0023}^{+0.0024}$ & $0.2359_{- 0.0039}^{+0.0042}$ & $0.2827_{- 0.0012}^{+0.0012}$ & $0.3693_{- 0.0033}^{+0.0034}$ & $0.2491_{- 0.0012}^{+0.0013}$ \\
$R_{\rm B}/R_{\rm A}$ & Radius Ratio & $0.12794_{- 0.00063}^{+0.00064}$ & $0.13362_{- 0.00041}^{+0.00041}$ & $0.16419_{- 0.00075}^{+0.0008}$ & $0.20446_{- 0.00037}^{+0.00035}$ & $0.23658_{- 0.001}^{+0.00098}$ \\
$S_{\rm B} / S_{\rm A}$ & Surface Brightness Ratio & $0.02375_{- 0.0018}^{+0.0018}$ & $0.04121_{- 0.0016}^{+0.0016}$  & $0.05141_{- 0.00085}^{+0.00083}$ & $0.07156_{- 0.00038}^{+0.00038}$ & $0.06839_{- 0.0012}^{+0.0012}$ \\
$\rho$ (g/cm$^3$) & Stellar Density & $0.3832_{- 0.009}^{+0.0089}$ & $0.298_{- 0.0061}^{+0.0062}$ & $0.378_{- 0.0039}^{+0.004}$ & $0.3084_{- 0.0016}^{+0.0017}$ & $1.206_{- 0.013}^{+0.014}$ \\
$T_{\rm A}$ (K) & Primary $T_{\rm eff}$ & 5880 $\pm150$ & 6450 $\pm120$ & 5807 $\pm78$ & 6390 $\pm 245$ & 5430 $\pm115$\\
$T_{\rm B}$ (K) & Secondary $T_{\rm eff}$ & 2672 $\pm52$ & 3090 $\pm38$ & 3026 $\pm26$ & 3340 $\pm78$ & 3055 $\pm 43$\\
\hline \multicolumn{7}{c}{Orbital Parameters} \\ \hline
$t_{\rm 0}$ (Days) & Time of Eclipse & $-1815.37339_{- 0.00043}^{+0.00043}$ & $4526.33509_{- 0.00041}^{+0.00043}$ & $5265.69871_{- 0.0002}^{+0.00018}$ & $8544.39821_{- 2.6e-05}^{+2.7e-05}$ & $-1553.31864_{- 0.00026}^{+0.00026}$ \\
$P_{\rm orb}$ (Days) & Orbital Period & $2.77868783_{-0.00000032}^{+0.00000033}$ & $5.54564705_{-0.00000047}^{+0.00000045}$ & $4.93655905_{-0.00000020}^{+0.00000021}$ & $5.147689048_{-0.000000099}^{+0.000000099}$ & $6.19197794_{-0.00000040}^{+0.00000040}$ \\
$b$ & Impact Parameter & $0.6382_{- 0.01012}^{+0.01014}$ & $0.36294_{- 0.01928}^{+0.01962}$ & $0.73196_{- 0.002758}^{+0.002657}$ & $0.63797_{- 0.001372}^{+0.001369}$ & $0.64393_{- 0.005239}^{+0.005063}$ \\
$e$ & Ecccentricity & $0.00199_{- 0.00051}^{+0.00054}$ & $0.19853_{-0.00014}^{+0.00014}$ & $0.121398_{-0.000072}^{+0.000074}$ & $0.000756_{-0.000048}^{+0.000048}$ & $0.12651_{- 0.00016}^{+0.00017}$ \\
$\omega$ $(^{\circ})$ & Argument of Periastron & $-2.1476_{- 0.64}^{+0.086}$ & $0.097908_{- 0.00096}^{+0.00096}$ & $-1.2724_{- 0.00034}^{+0.00033}$ & $1.7221_{- 0.035}^{+0.033}$ & $-1.3544_{- 0.00067}^{+0.00069}$ \\
$\gamma$ (km/s) & Systemic Velocity & $19.3553030_{-0.0000099}^{+0.0000099}$ & $4.15_{-0.060}^{+0.061}$ & $28.9561_{-0.0060}^{+0.0061}$ & $28.0391_{-0.0038}^{+0.0037}$ & $-48.015684_{-0.000098}^{+0.000098}$ \\
$\dot{\gamma}$ & Linear Trend (m/s/year) & $57.7_{-2.2}^{+2.2}$ & $-3.65_{-3.6}^{+3.6}$ \\
$I$ $(^{\circ})$ & Inclination & $83.52_{- 0.15}^{+0.15}$ & $87.32_{- 0.16}^{+0.16}$ & $85.499_{- 0.029}^{+0.029}$ & $85.548_{- 0.013}^{+0.013}$ & $87.720_{- 0.021}^{+0.021}$\\
$a$ ($R_{\odot}$) & Semi-major Axis & $9.226_{- 0.024}^{+0.024}$ & $14.68_{- 0.22}^{+0.23}$ & $14.41_{- 0.033}^{+0.033}$ & $14.86_{- 0.13}^{+0.14}$ & $15.18_{- 0.047}^{+0.048}$ \\
$a$ (AU) & Semi-major Axis & $0.04291_{-0.00011}^{0.00011}$ & $0.06829_{-0.001}^{0.0011}$ & $0.06701_{-0.00015}^{0.00015}$ & $0.0691_{-0.00062}^{0.00064}$ & $0.07061_{-0.00022}^{0.00022}$ \\
\hline \multicolumn{7}{c}{RM Parameters} \\ \hline
$v \sin I$ (km/s) & Projected Rotational Velocity & $31.61_{- 0.47}^{+0.47}$ & $17.25_{- 0.89}^{+0.89}$ & $16.54_{- 0.52}^{+0.54}$ & $16.32_{- 0.29}^{+0.28}$ & $7.83_{- 0.29}^{+0.28}$ \\
$|\lambda|$ $(^{\circ})$ & Sky-Projected Obliquity & $4.3_{- 2.1}^{+2.1}$ & $12.0_{-5.1}^{+4.9}$ & $2.7_{- 1.2}^{+1.3}$ & $0.31_{- 0.71}^{+0.71}$ & $3.3_{- 1.9}^{+1.9}$ \\
$\psi$ $(^{\circ})$ & True Obliquity & $7.9_{- 1.1}^{+1.2}$ & $13.6_{- 5.3}^{+5.4}$ & $16.5_{-6.3}^{+4.7}$ & $18.0_{-5.2}^{+4.1}$ & $17.3_{-7.2}^{+5.3}$ \\
$\zeta$ (km/s) & Macroturbulent Velocity & $4.3_{- 2.5}^{+2.2}$ & $3.9_{- 2.5}^{+2.3}$ & $4.8_{- 1.5}^{+1.3}$ & $14.4_{- 1.1}^{+1.1}$ & $3.6_{- 2.3}^{+1.9}$ \\
$\beta$ (km/s) & Thermal Velocity & $1.05_{- 0.7}^{+0.65}$ & $0.504_{- 0.34}^{+0.34}$ & $1.02_{- 0.35}^{+0.33}$ & $1.03_{- 0.35}^{+0.34}$ & $2.07_{- 1.5}^{+1.4}$ \\
$\gamma$ (km/s) & Lorentzian Velocity & $3.79_{- 0.58}^{+0.55}$ & $3.83_{- 0.92}^{+0.87}$ & $4.28_{- 0.62}^{+0.58}$ & $4.73_{- 0.23}^{+0.21}$ & $0.886_{- 0.71}^{+0.73}$ \\
$P_{\rm rot}$ (days) & Rotational Period & $2.85_{- 0.12}^{+0.12}$ & $5.38_{- 0.04}^{+0.04}$ & $4.93_{- 0.066}^{+0.065}$ & $5.18_{- 0.13}^{+0.13}$ & $6.43_{- 0.079}^{+0.079}$ \\
$\tau_{\rm \psi}$ (yr) & Realignment Timescale & $2.58 \times 10^5$ & $6.40 \times 10^7$ & $3.60 \times 10^6$ & $2.20 \times 10^7$ & $1.62 \times 10^7$\\
$\tau_{\rm sync}$ (yr) & Synchronization Timescale & $1.29 \times 10^5$ & $3.20 \times 10^7$ & $1.80 \times 10^6$ & $1.10 \times 10^7$ & $8.10 \times 10^6$\\
$\tau_{\rm circ, A}$ (yr) & Circularization Timescale (Primary) & $8.97 \times 10^5$ & $9.33 \times 10^8$ & $3.52 \times 10^7$ & $4.47 \times 10^8$ & $3.25 \times 10^8$\\
$\tau_{\rm circ, B}$ (yr) & Circularization Timescale (Secondary) & $1.08 \times 10^8$ & $3.64 \times 10^9$ & $8.28 \times 10^8$ & $3.81 \times 10^8$ & $5.32 \times 10^9$\\ 
\\
\hline
\end{tabular}
\end{table*}

\bibliographystyle{mnras}
\bibliography{EBLMIX} % if your bibtex file is called example.bib

%%%%%%%%%%%%%%%%%%%%%%%%%%%%%%%%%%%%%%%%%%%%%%%%%%

\renewcommand{\arraystretch}{1.5}

\appendix

\section{RV, Photometry, and Corner Plots}

\begin{figure*}
  \centering
  \begin{tabular}{cc}
    \includegraphics[width=0.49\textwidth]{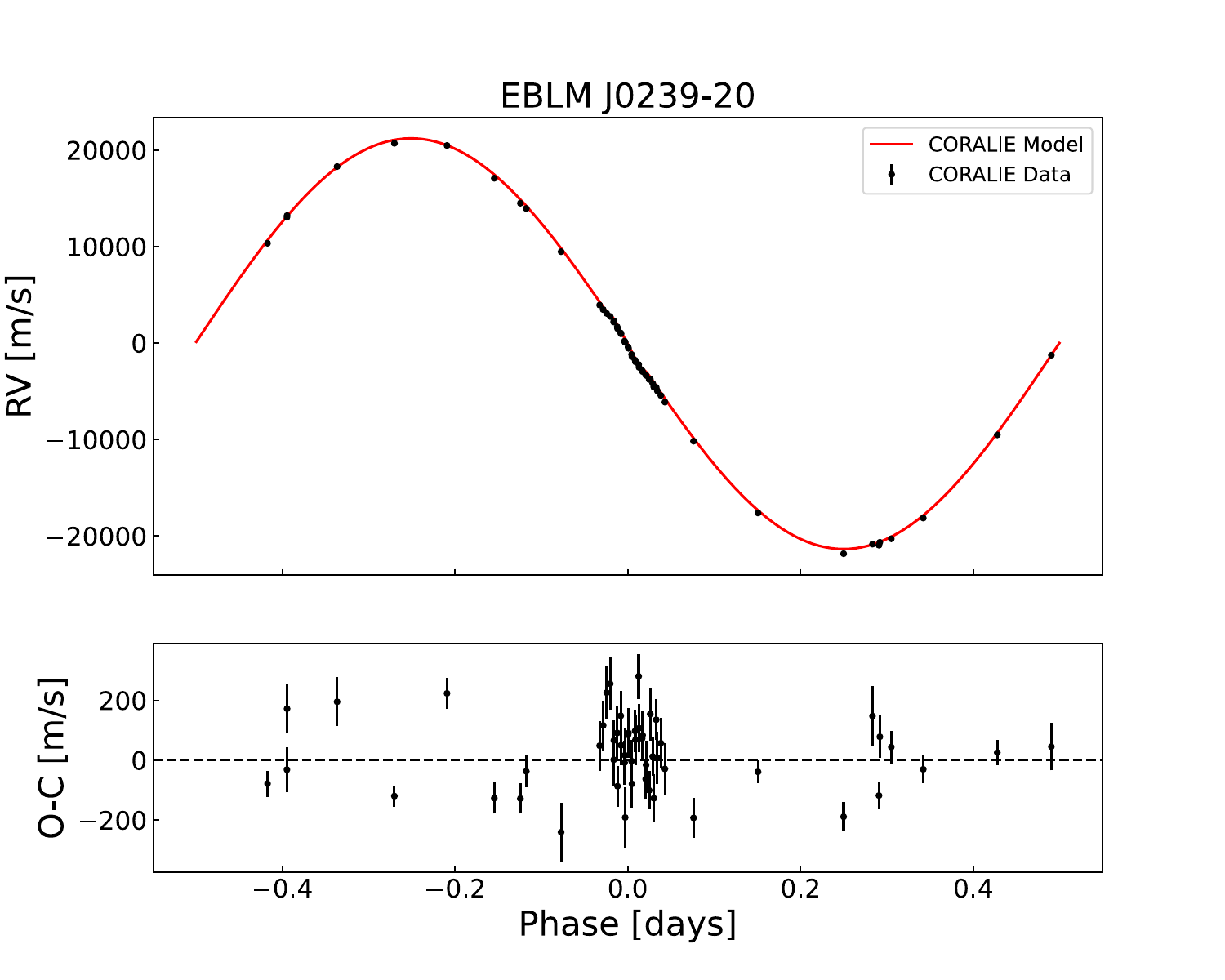} &
    \includegraphics[width=0.49\textwidth]{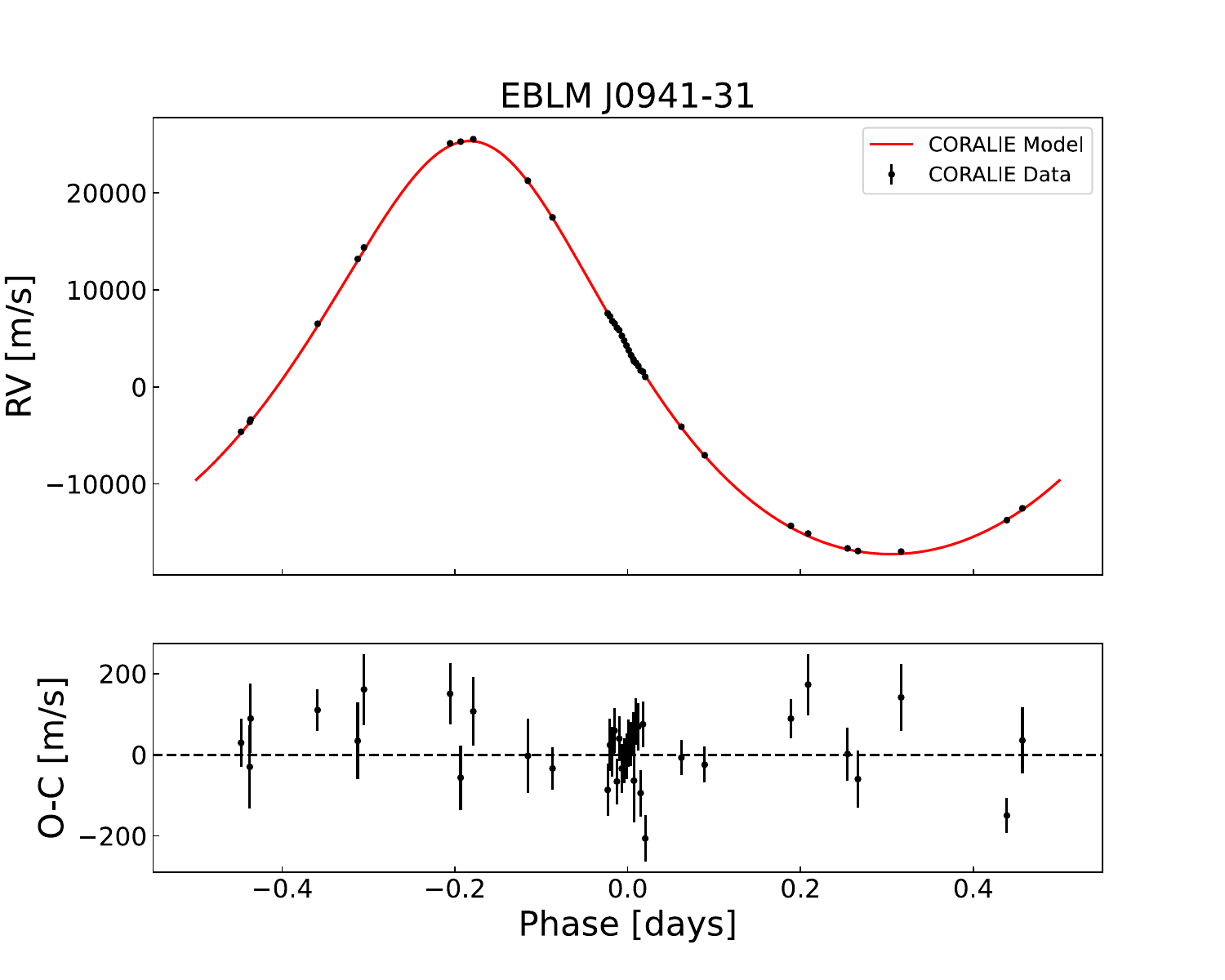} \\
    
    \includegraphics[width=0.49\textwidth]{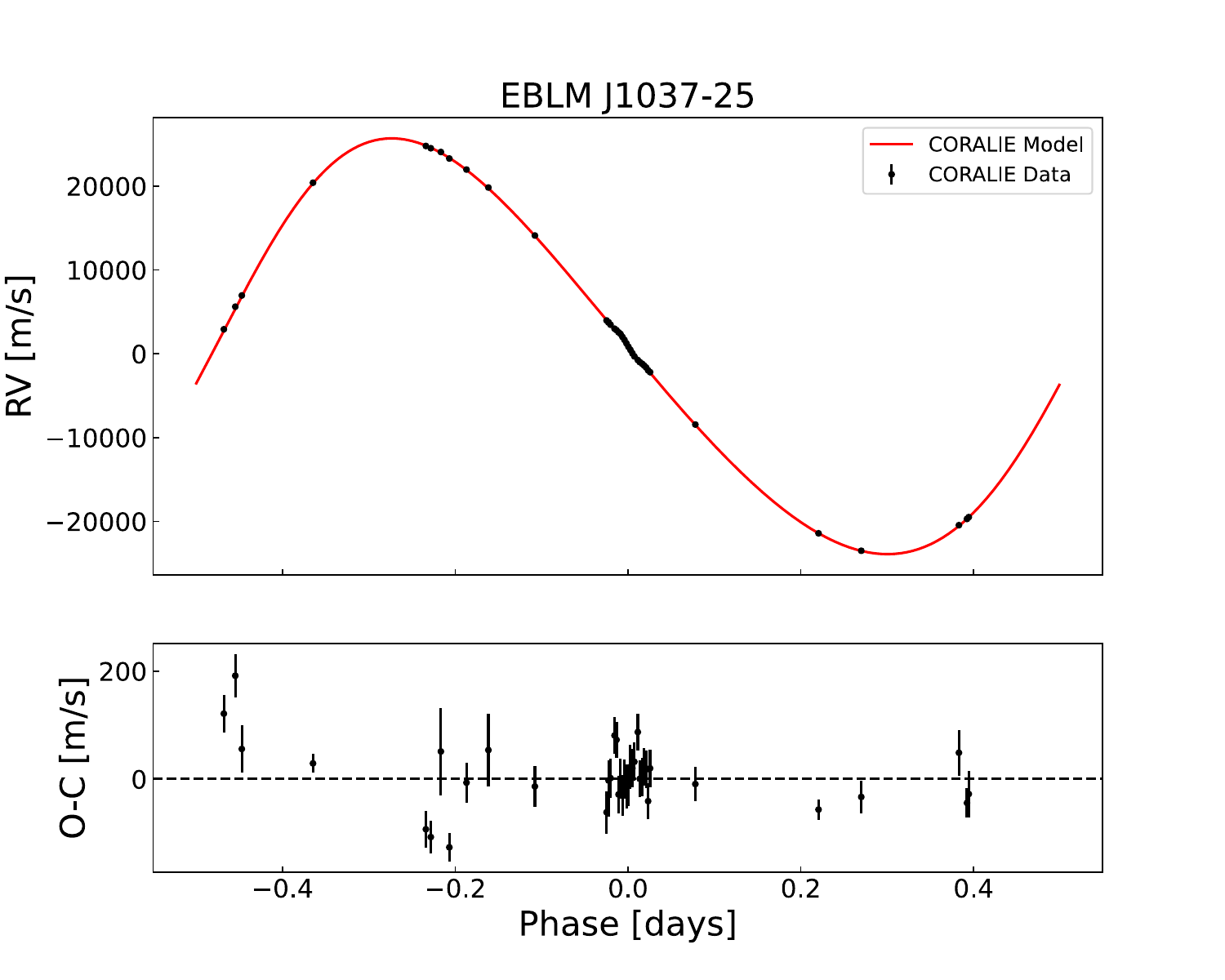} &
    \includegraphics[width=0.49\textwidth]{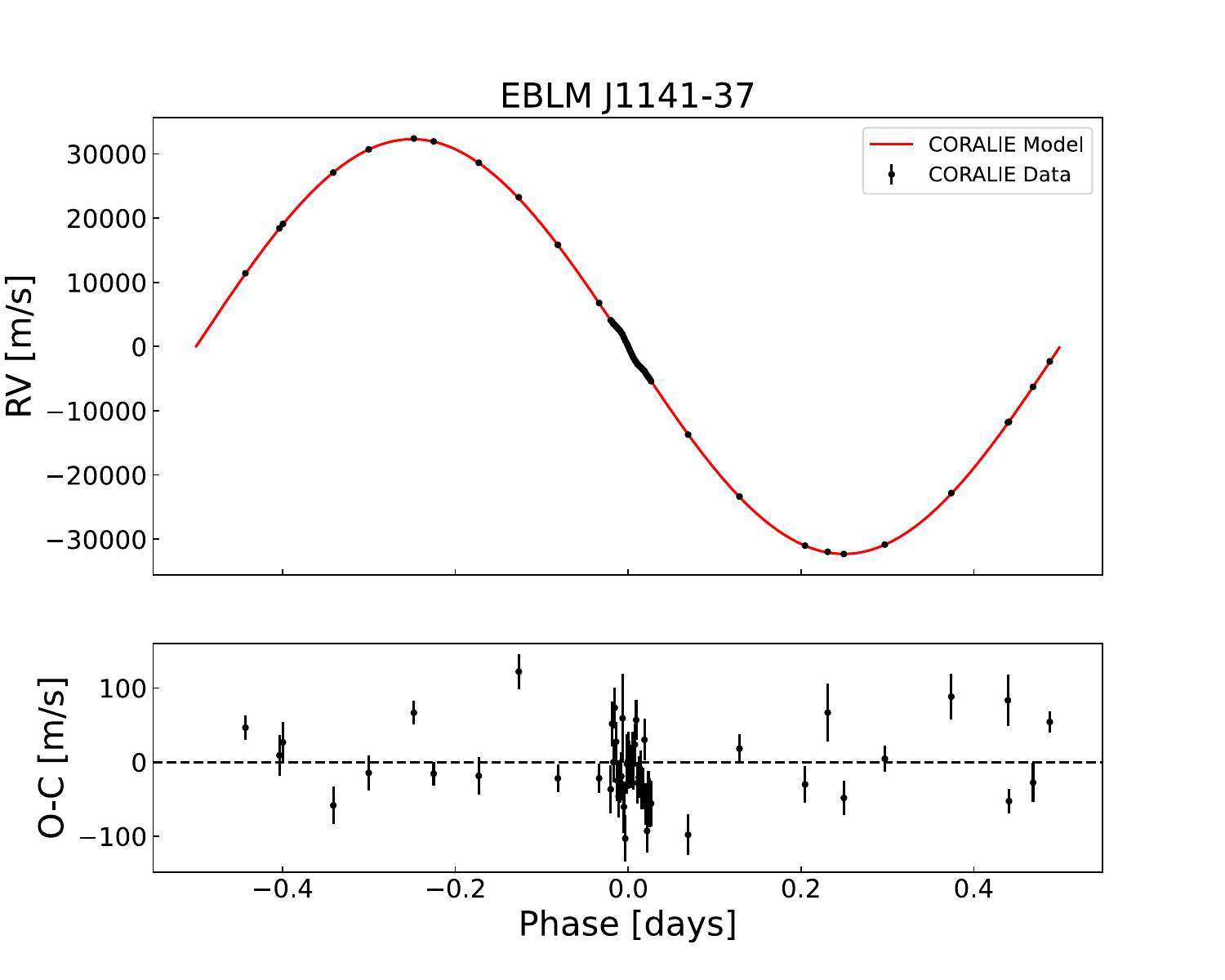} \\
    
    \includegraphics[width=0.49\textwidth]{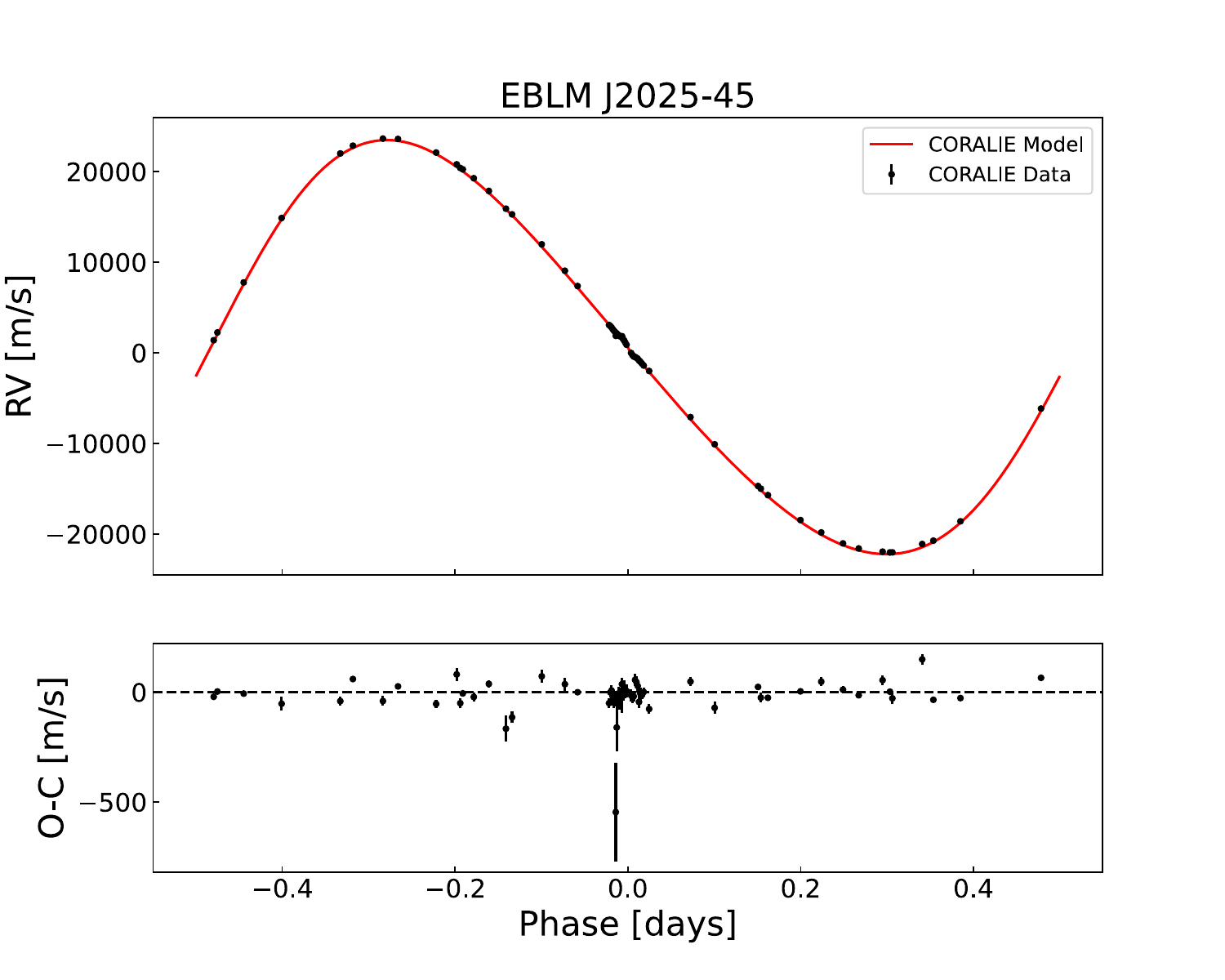} &
    \parbox[b][0.35\textwidth][c]{0.49\textwidth}{%
      % \centering
      \caption{\coralie ~Radial velocity observations of each system, phase-folded on the best fitting orbital period. The data are shown by the black dots, with error bars on the order of m/s and are invisible at this scale. The red line shows the {\tt \textsc{PyMC3}} best fit model \textit{with} the RM fit as well. The RM here is only slightly visible because it happens on the order of m/s compared to the Keplerian, which is on the order of km/s.}\label{fig:RVs}}
  \end{tabular}

\end{figure*}

\begin{figure*}
\centering
  \begin{tabular}{cc}
    \includegraphics[width=0.49\textwidth]{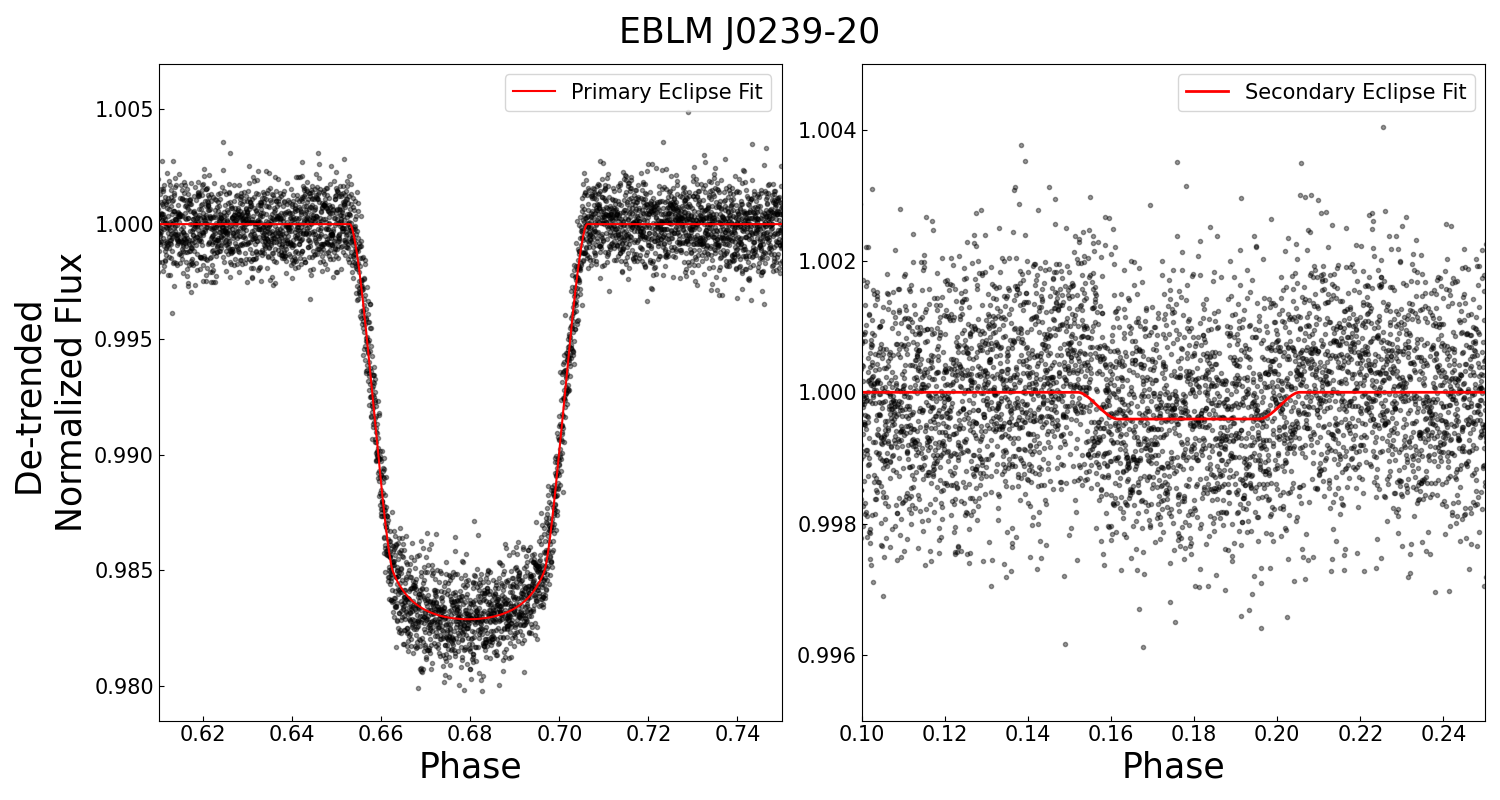} &
    \includegraphics[width=0.49\textwidth]{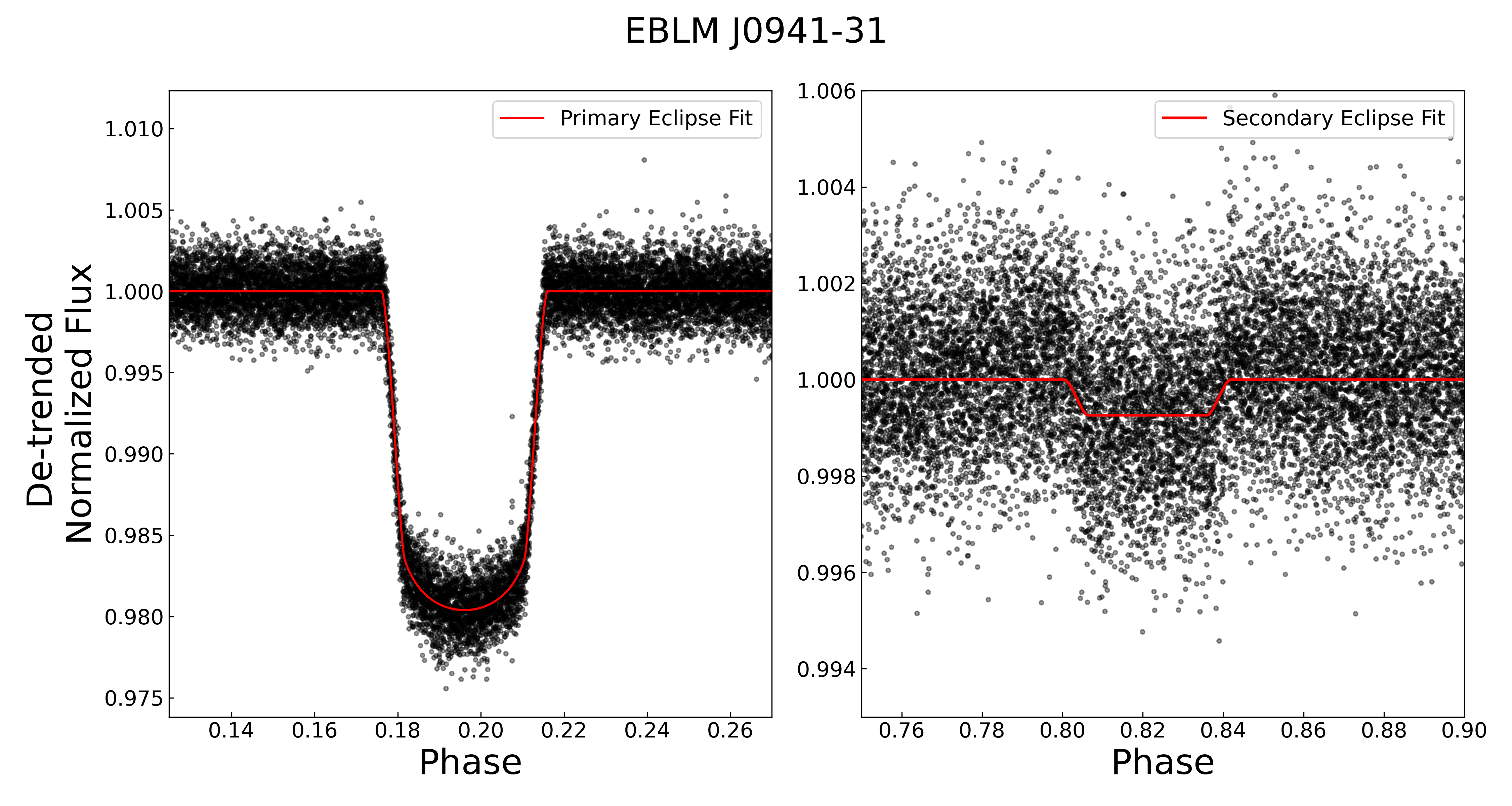} \\
    
    \includegraphics[width=0.49\textwidth]{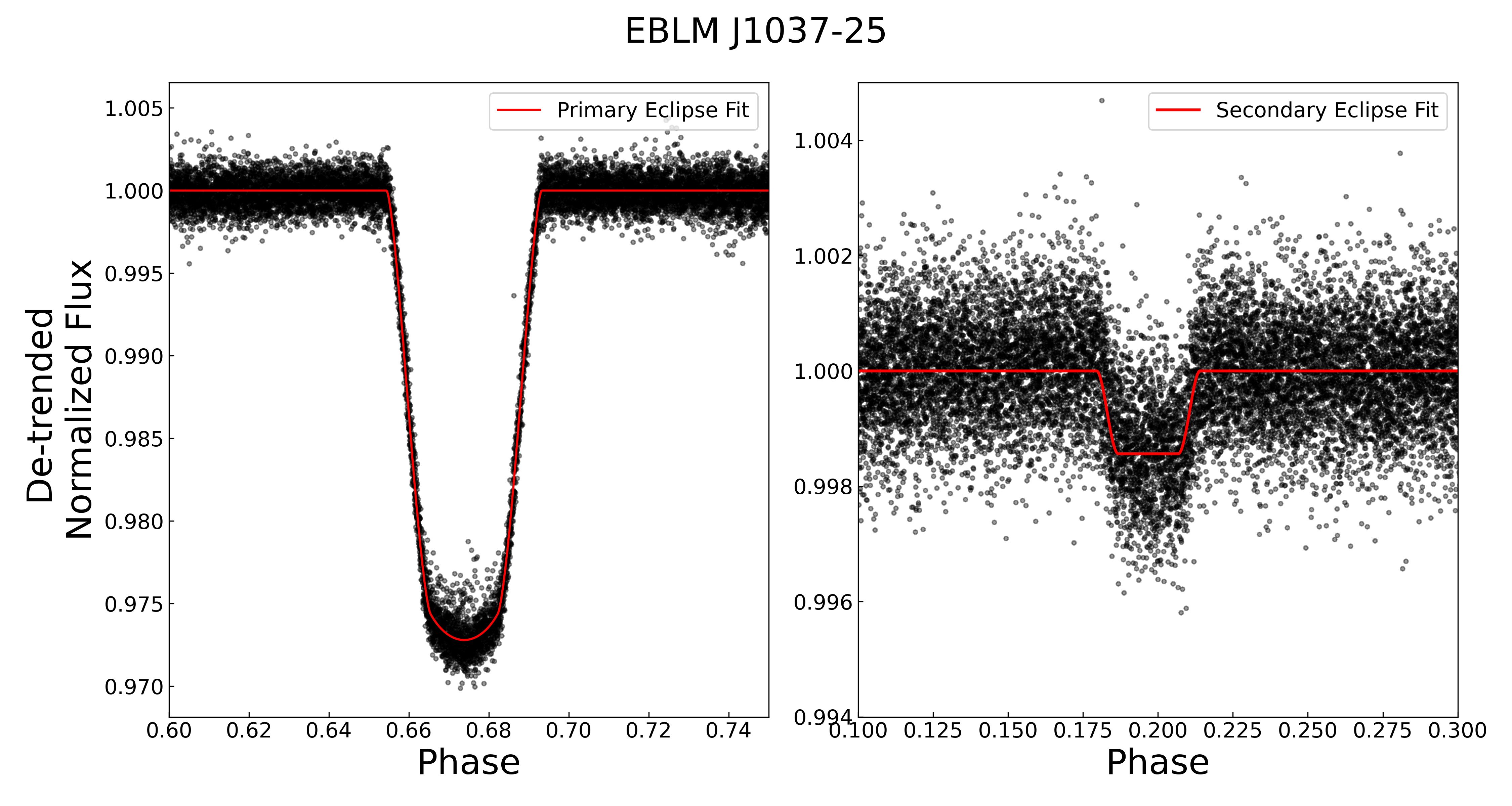} &
    \includegraphics[width=0.49\textwidth]{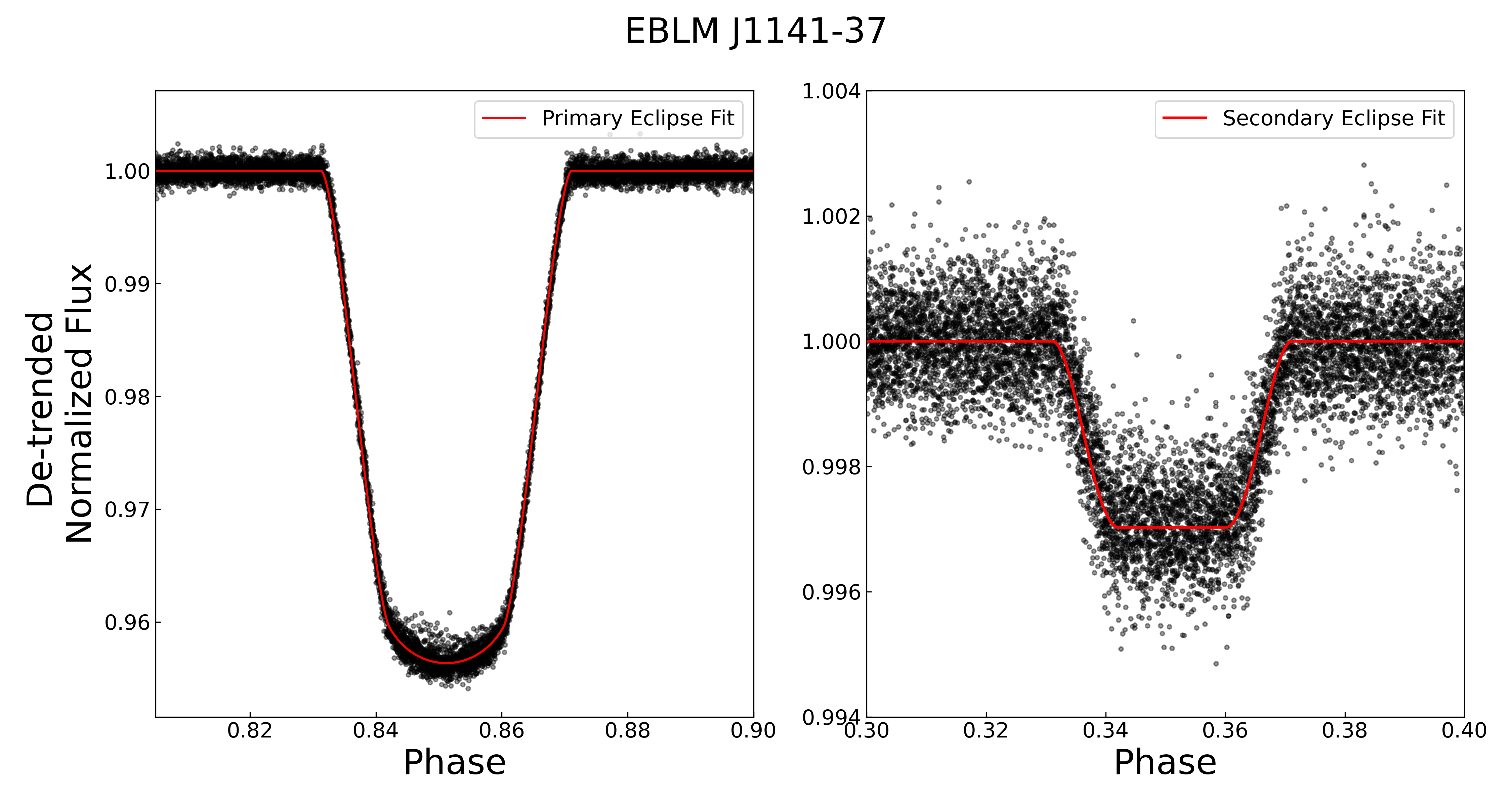} \\
    
    \includegraphics[width=0.49\textwidth]{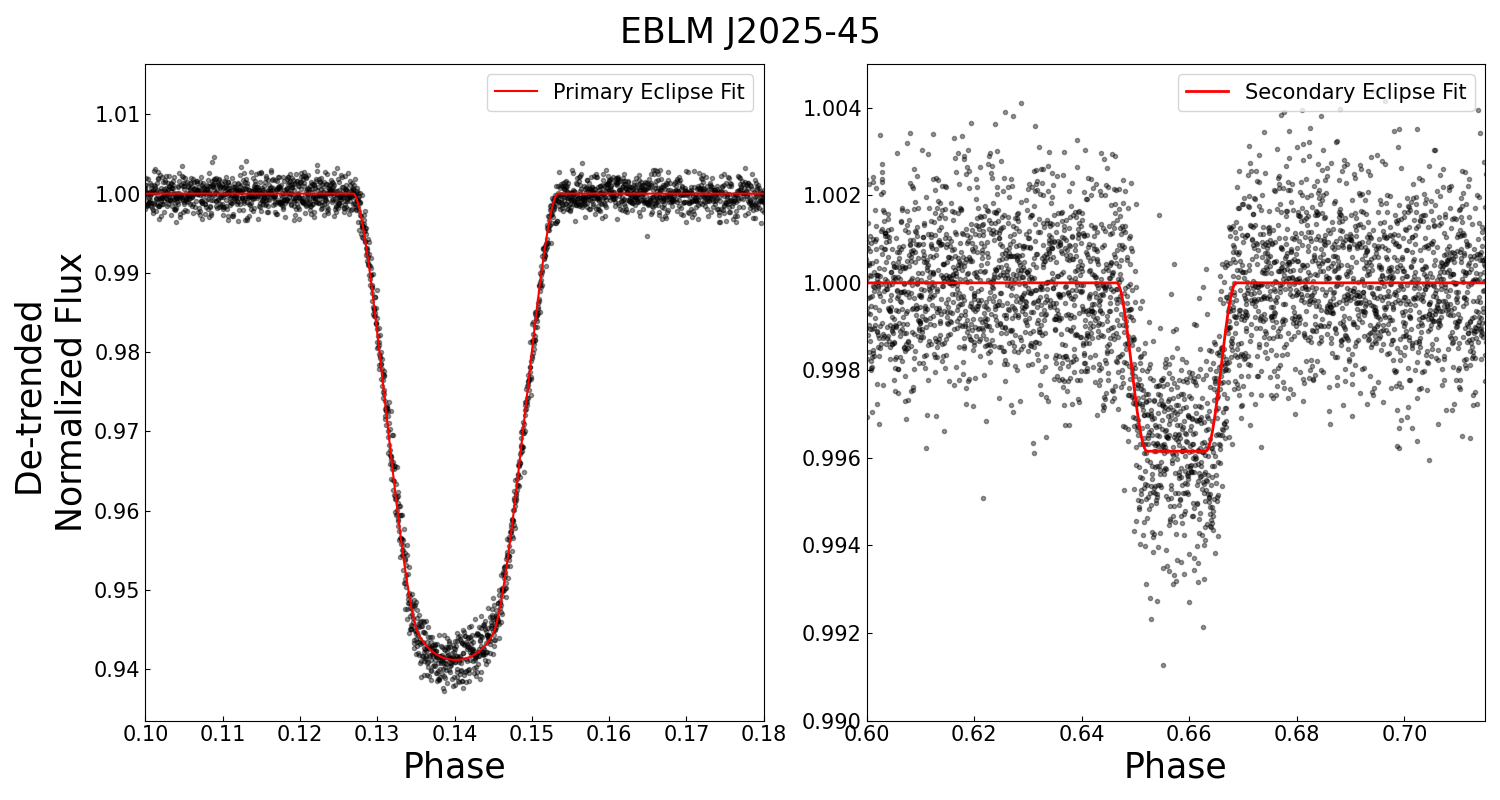} &
    \parbox[b][0.25\textwidth][c]{0.49\textwidth}{%
      % \centering
      \caption{Primary and secondary eclipse plots for each of the systems. The light curves were phase-folded over the orbital period found from the MCMC. The red line shows the MCMC fit.}\label{fig: Eclipses and Light curves}}
  \end{tabular}

\end{figure*}

\begin{figure*}
    \includegraphics[width=0.99\textwidth]{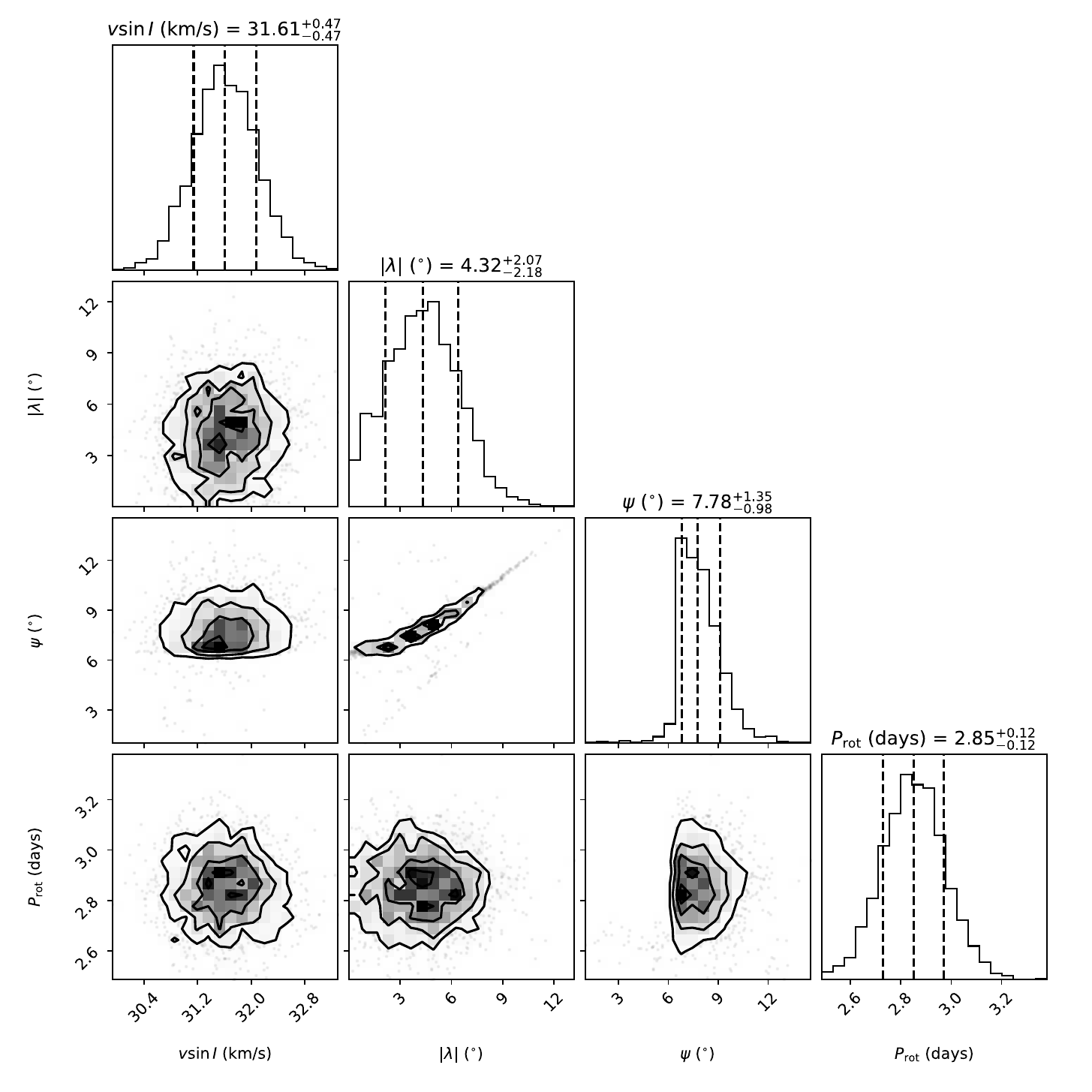}
    \caption{Corner plot for EBLM J0239-20 produced from the MCMC simulation of the joint fit to all RVs (Keplerian and Rossiter) and the TESS photometry for primary and secondary eclipses. The plot shows the correlation between the parameters corresponding to the RM fit - $v_{\rm rot} \sin I_{\rm \star}$, $\lambda$, $\psi$, \& $P_{\rm rot}$.}\label{fig: Corner Plot 1}
\end{figure*}

\begin{figure*}
    \includegraphics[width=0.99\textwidth]{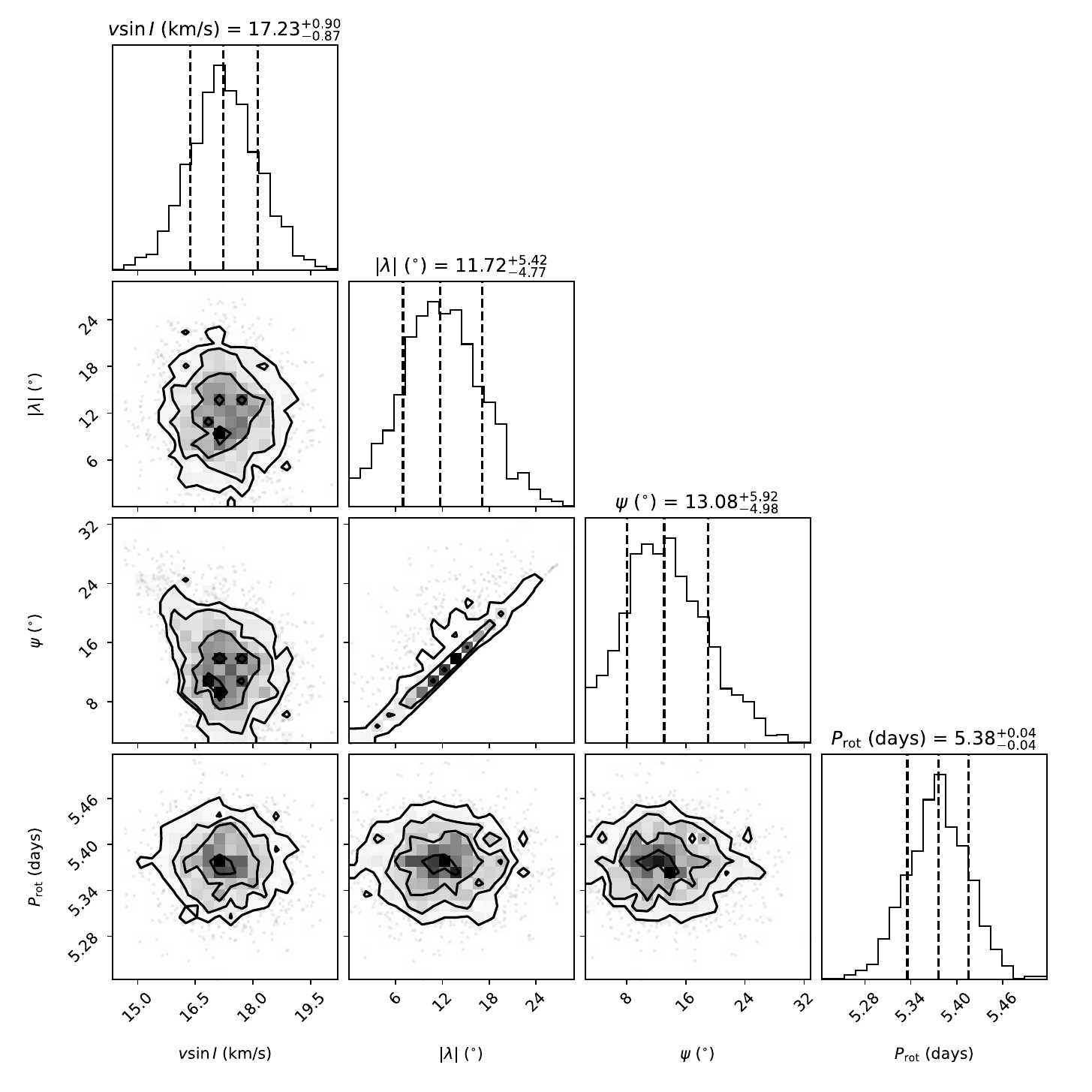}
    \caption{Corner plot for EBLM J0941-31 produced from the MCMC simulation of the joint fit to all RVs (Keplerian and Rossiter) and the TESS photometry for primary and secondary eclipses. The plot shows the correlation between the parameters corresponding to the RM fit - $v_{\rm rot} \sin I_{\rm \star}$, $\lambda$, $\psi$, \& $P_{\rm rot}$.}\label{fig: Corner Plot 2}
\end{figure*}

\begin{figure*}
    \includegraphics[width=0.99\textwidth]{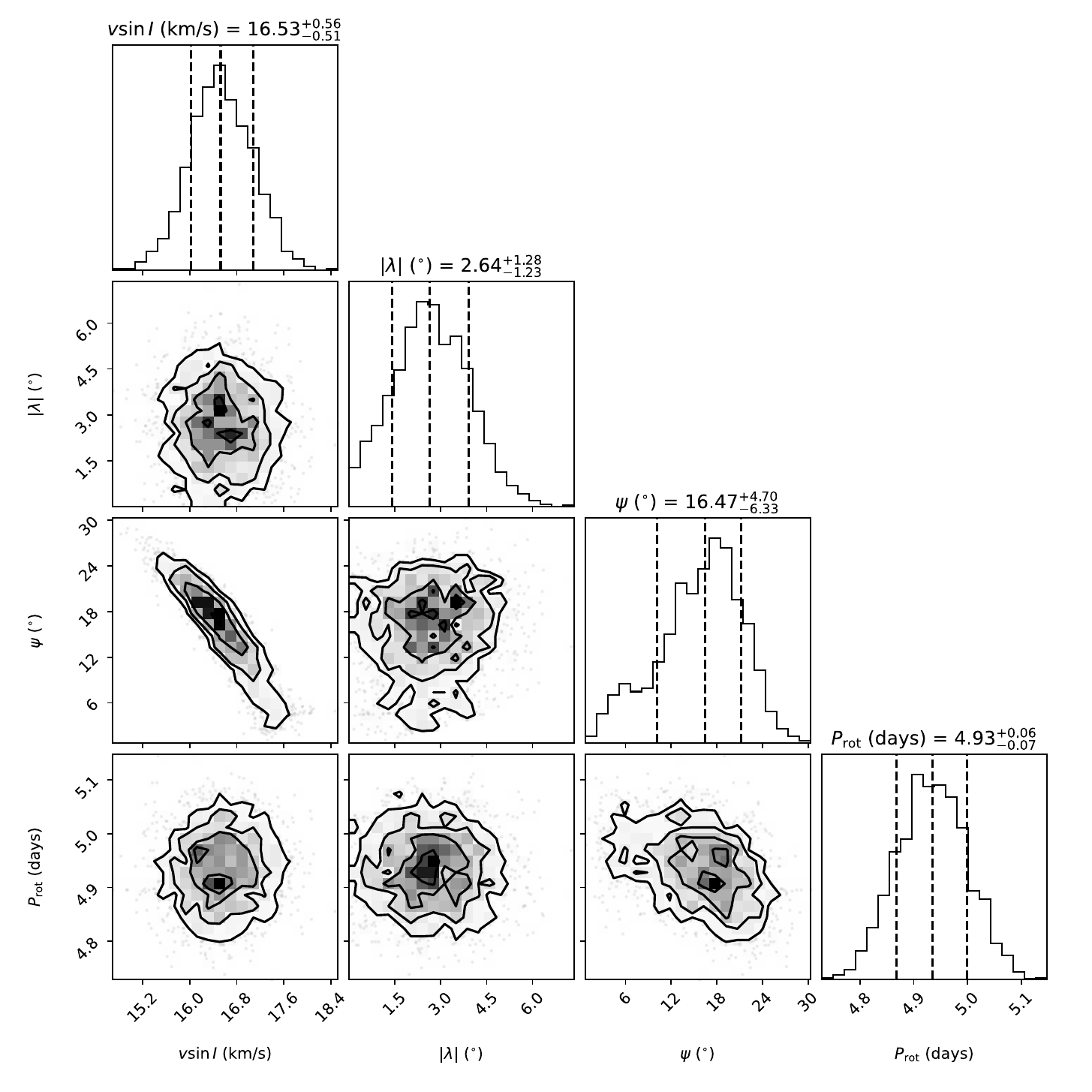}
    \caption{Corner plot for EBLM J1037-25 produced from the MCMC simulation of the joint fit to all RVs (Keplerian and Rossiter) and the TESS photometry for primary and secondary eclipses. The plot shows the correlation between the parameters corresponding to the RM fit - $v_{\rm rot} \sin I_{\rm \star}$, $\lambda$, $\psi$, \& $P_{\rm rot}$.}\label{fig: Corner Plot 3}
\end{figure*}

\begin{figure*}
    \includegraphics[width=0.99\textwidth]{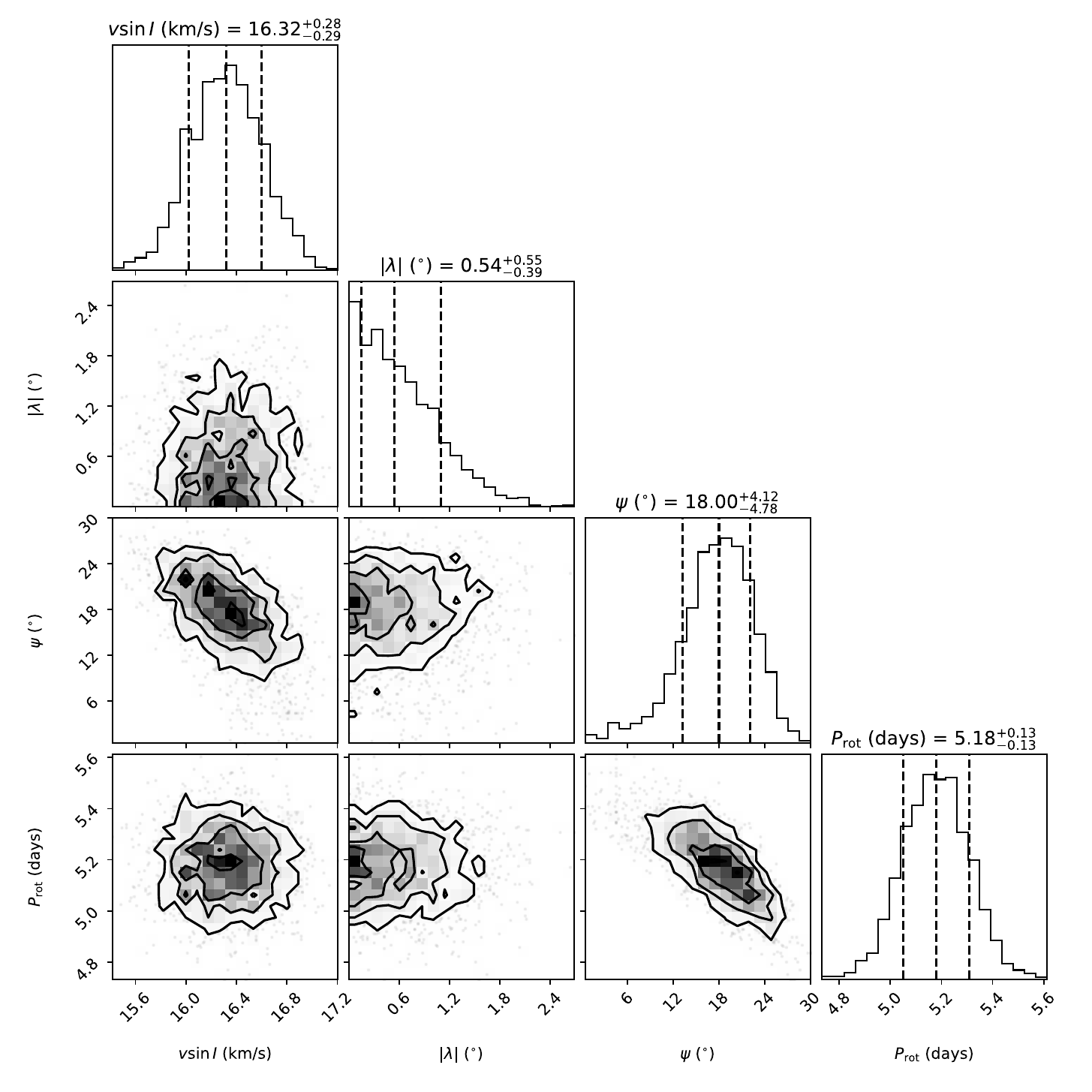}
    \caption{Corner plot for EBLM J1141-37 produced from the MCMC simulation of the joint fit to all RVs (Keplerian and Rossiter) and the TESS photometry for primary and secondary eclipses. The plot shows the correlation between the parameters corresponding to the RM fit - $v_{\rm rot} \sin I_{\rm \star}$, $\lambda$, $\psi$, \& $P_{\rm rot}$.}\label{fig: Corner Plot 4}
\end{figure*}

\begin{figure*}
    \includegraphics[width=0.99\textwidth]{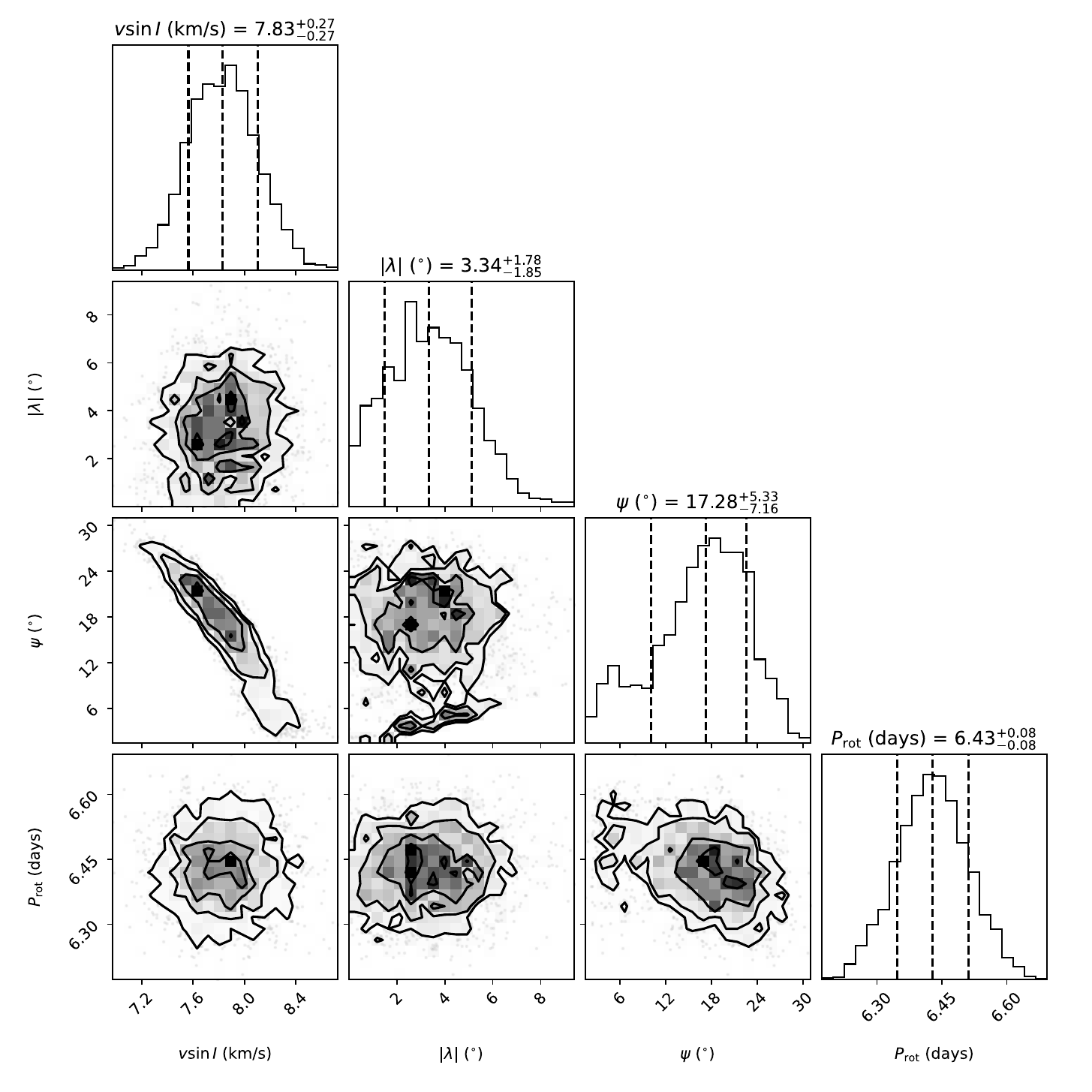}
    \caption{Corner plot for EBLM J2025-45 produced from the MCMC simulation of the joint fit to all RVs (Keplerian and Rossiter) and the TESS photometry for primary and secondary eclipses. The plot shows the correlation between the parameters corresponding to the RM fit - $v_{\rm rot} \sin I_{\rm \star}$, $\lambda$, $\psi$, \& $P_{\rm rot}$.}\label{fig: Corner Plot 5}
\end{figure*}

\newpage

% Don't change these lines
\bsp	% typesetting comment
\label{lastpage}
\end{document}